\documentclass[5p]{elsarticle}

\usepackage{soul}
\usepackage{booktabs}
\usepackage{multirow}
\usepackage{aas_macros}
\usepackage{amsmath}
\usepackage{xcolor}
\usepackage{hyperref}

\journal{Astronomy and Computing}

\biboptions{authoryear}

\begin{document}

\begin{frontmatter}

\title{A proof-of-concept neural network for inferring parameters of a black hole \\ from partial interferometric images of its shadow}

\author[mysecondaryaddress]{Anton A. Popov\fnref{popovfn}}
\fntext[popovfn,surdyaevfn]{Affiliations at the time this work was completed.}

\author[mymainaddress]{Vladimir Strokov\corref{mycorrespondingauthor}}
\cortext[mycorrespondingauthor]{Corresponding author}
\ead{vstroko1@jhu.edu}

\author[mysecondaryaddress]{Aleksey A. Surdyaev\fnref{surdyaevfn}}

\address[mymainaddress]{Department of Physics and Astronomy, Johns Hopkins University, Baltimore, MD 21218 USA}
\address[mysecondaryaddress]{Lebedev Physical Institute, Astro Space Centre,
84/32 ul. Profsoyuznaya, Moscow, Russia 117997}

\begin{abstract}
We test the possibility of using a convolutional neural network to infer the inclination angle of a black hole \textit{directly} from the incomplete image of the black hole's shadow in the $uv$-plane. To this end, we develop a proof-of-concept network and use it to explicitly find how the error depends on the degree of coverage, type of input and coverage pattern. We arrive at a typical error of $10^\circ$ at a level of absolute coverage $1\%$ (for a pattern covering a central part of the $uv$-plane), $0.3\%$ (pattern covering the central part and the periphery, the $0.3\%$ referring to the central part only), and $14\%$ (uniform pattern). These numbers refer to a network that takes both amplitude and phase of the visibility function as inputs. We find that this type of network works best in terms of the error itself and its distribution for different angles. In addition, the same type of network demonstrates similarly good performance on highly blurred images mimicking sources nearing being unresolved. In terms of coverage, the magnitude of the error does not change much as one goes from the central pattern to the uniform one. We argue that this may be due to the presence of a typical scale which can be mostly learned by the network from the central part alone.
\end{abstract}


\end{frontmatter}

\section{Introduction}\label{intro}

There is vast indirect evidence of massive compact objects residing in galactic centres~\citep{BHmass_determination,Genzel_2010}. Measurements of the masses of these objects~\citep[see also][]{McConnell_2012} and their compact sizes plausibly suggest that they are supermassive black holes. If so, the light coming from surrounding matter must be strongly lensed to form distinctive silhouettes of the black holes~\citep[][and references therein]{Luminet,luminet_review}. 

The dedicated Event Horizon Telescope (EHT) array \citep{EHT-3,EHT-2} had been resolving increasingly closer neighborhoods of Sgr~A$^\star$ and~M87$^\star$ \citep{Doeleman_SgrA, 1_3mmVLBI_SgrA, Doeleman_2012, Akiyama_2015, Lu_2018} before these efforts culminated in the historic first direct image of a black hole~\citep{EHTI,EHTII,EHTIII,EHTIV,EHTV,EHTVI}. Also contributing to the task of black hole imaging are other existing arrays~\citep[e.g.][and references therein]{Issaoun_2019,Brinkerink_2016} and upcoming projects \citep{Kardashev_2014,Ivanov_2019,Hong_2014,%
Hong_2019}.

It is known~\citep{Thompson_2017} that a network of very-long-baseline interferometry (VLBI) stations, such as EHT, aims at measuring complex-valued visibility function
\begin{equation}
\mathcal{V}(u,v) = \iint{e^{-2\pi i(u\alpha + v\beta)}}I(\alpha,\beta)\,d\alpha d\beta\,, 
\end{equation}
which is the spatial Fourier transform of brightness distribution~$I$ in image plane $(\alpha,\beta)$. If the angles $\alpha$ and $\beta$ are given in units of characteristic angular resolution $\lambda/D$, then spatial frequencies in the $uv$-plane are measured in units of $D/\lambda$, where $D$ is a characteristic baseline and $\lambda$, the working wavelength. Since, in reality, the coverage of the $uv$-plane is always partial, the EHT team used a variety of techniques to reconstruct the image of the black hole shadow, such as the CLEAN algorithm and regularized maximum likelihood methods~\citep[e.g.][and references therein]{CLEAN-1,CLEAN-2,Narayan_1986,EHTIV}.

On the other hand, another set of algorithms known as convolutional neural networks (CNNs) has proven to be extremely effective in the general problem of image recognition~\citep{russakovsky,keras}. In recent years, (artificial) neural networks in general and CNNs in particular have been finding more and more applications in astrophysics: to name a few, automated analysis and detection of strong gravitational lenses~\citep{Hezaveh_2017,Petrillo_2017,Jacobs_2017,Lanusse_2017,%
Pourrahmani_2018,Schaefer_2018}, dark matter halo simulations~\citep{Agarwal_2018,Berger_2018}, black hole identification in globular clusters~\citep{Askar_2019}, and computing the mass of forming planets~\citep{Alibert_2019}. Recurrent Inference Machines~\citep{Putzky_2017} were also used to process interferometric observations of strong lenses~\citep{Morningstar_2018,Morningstar_2019} as well as found applications in medical imaging~\citep{Loenning_2019}.

Also, \citet{deep_horizon} developed two convolutional networks called Deep Horizon to recover accretion and black hole parameters from real-space images. However, as mentioned, VLBI observations rather yield partial Fourier transform of the images, and it would be more natural if a neural network took the Fourier image directly as its input. In that case, it is crucial to investigate how the error of the output depends on the degree of coverage of the $uv$-plane. 

There are a few reasons to use a neural network to infer parameters of a black hole. The first is the speed of analysis. For a given image and a given coverage, parameter inference algorithms such as MCMC~\citep{MCMC_Broderick} take a lot of time as they randomly walk in the parameter space, and, for a different coverage, this process should be repeated from the very beginning. This becomes especially important in the case of black hole silhouettes when generating one at each step of the random walk is time-consuming, because typically ray-tracing is used. A neural network, on the other hand, needs to be trained only once and on a set of images which is generated once and for all. Another reason is that predictions of a neural network can be used to double check the values of the parameters obtained with other techniques. Finally, a neural network could be integrated into a parameter inference pipeline. For example, below, we train our networks to operate within a wide range of degrees of coverage. This approach not only saves time (we need to train them once rather than training a series of networks on its own degree of coverage each) but also makes the networks more universal. Then, such a network could be used in the first stage of a pipeline to determine ranges of the parameters which can be further narrowed down with traditional methods.

In this Note we develop a convolutional neural network that determines inclination angle\,\footnote{This is the angle between the black hole's spin and the line of sight of a distant observer}~$\theta$ from a partially covered image of the $uv$-plane. Our aim is to study how the performance of the network depends on the degree of coverage, types of input and different coverage patterns that emulate different observational settings.

In more detail, firstly, following~\citet{Hezaveh_2017}, we estimate the error introduced by the CNN by the width of the deviation distribution. We also check whether the distribution is Gaussian and study how the error depends on the degree of the $uv$-plane coverage. Secondly, we evaluate different types of input: the amplitude of the visibility function alone, the amplitude and the phase, and the same options accompanied by a mask that encodes which pixels contain a signal. Finally, the network is fed with different patterns of the $uv$-plane coverage: a) only a central part of the plane is covered (which mimics the case of EHT), b) in addition to the central part, there is a covered ring a few times bigger than the center (which mimics the case of future space interferometry projects with antennas in high orbits, \citet{Fish_2019}), and c) uniform coverage (reminiscent of the setting described in~\citep{Palumbo}). We train four networks that differ in their inputs on the three coverage patterns each. Regarding the degree of coverage, training, validation and test sets include all levels of coverage (in a certain range), which implies that the networks were trained to do a prediction with an arbitrary number of activated pixels (if they follow one of the three patterns).

Since the accessible region of the $uv$-plane in pattern (c) is larger than that of patterns (b) and, especially, (a), we will present our final results in terms of \textit{absolute} coverage. For patterns (a) and (b), we define it as the number of activated pixels in the central part of a Fourier image divided by the total number of pixels in pattern (c). For pattern (c) we define the coverage as the total number of activated pixels by the size of an image. At the same time, we will be presenting our intermediate results (Figs.~\ref{hist_mask=False_phase=False}--%
\ref{hist_mask=True_phase=True}) in terms of \textit{relative} coverage for pattens (a) and (b), where we define it w.r.t. the size of the central partand assume that the area of the latter constitutes $2.5\%$ of the total area accessible in pattern (c). We describe this point in more details in Subsect.~\ref{dataset}.

The structure of the Note is as follows. In the next section we describe our method while Sect.~\ref{results} contains comparative results for the cases described. There we also  discuss the prospects of improving the network to suit the real-observation needs.

\section{Method}\label{architecture}

The general problem of fitting data is finding an approximate mapping between observation(s)~$\mathbf{x}$ (e.g. the frequency at which the black-body radiation peaks) and an inferred value(s)~$\mathbf{y}$ (the black-body temperature), $\mathbf{x}\rightarrow\mathbf{y}$. To this effect, a hypothetical mapping which depends on parameters~$\mathbf{w}$ is introduced (in the simplest case of the linear hypothesis, there are only 2 parameters). Then, the parameters~$\mathbf{w}$ are adjusted so that the cumulative error (for example, the sum of squared deviations between the values predicted by the hypothesis and the actual values) is minimal.

Neural networks are a wide class of algorithms used to implement hightly nonlinear mappings~\citep{bishop}. For instance, $\mathbf{x}$ could be an image represented as a matrix of pixels and $\mathbf{y}$, say, the type of object in that image represented as a vector of logical ones and zeros (e.g. the 1st component encodes object ``human'' and will be equal to one if there is actually a human in the image and zero otherwise). Typically, a neural network is organized in a sequence of layers where each layer has its own parameters $\mathbf{w}_1, \mathbf{w}_2, \ldots$\,. The forward pass that maps $\mathbf{x}$ (input layer) to $\mathbf{y}$ (output layer) starts from composition $\mathbf{w}_1\circ \mathbf{x}$ (most often, it is the matrix multiplication) being fed to the so-called activation function $\mathbf{g}_2$ of the 2nd layer. The composition of the result with parameters~$\mathbf{w}_2$, $\mathbf{w}_2\circ \mathbf{g}_2\left(\mathbf{w}_1\circ \mathbf{x}\right)$, is fed to the activation function of the 3rd layer, and so on up to the last layer whose output is $\mathbf{y}$. The activation functions are required to be non-linear. The universal approximation theorem~\citep[e.g.][]{Cybenko_1989} states that such a network with a single hidden layer can approximate continuous functions arbitrary well, provided that the number of neurons in the layer (the dimension of $\mathbf{w}_1\circ \mathbf{x}$) is sufficient.

The process of adjusting weights of a neural network is known as training, which is achieved through minimizing the error of the predicted values~$\mathbf{y}_{\rm pred}$ on a training set~$\left\{\mathbf{x}_{\rm train}, \mathbf{y}_{\rm train}\right\}$ (for example, with the gradient-descent algorithm). The performance of the network during training is monitored by its error on a separate validation set~$\left\{\mathbf{x}_{\rm val}, \mathbf{y}_{\rm val}\right\}$. The final error is evaluated on a test set unavailable to the network during the training and measures the ability of the network to generalize to new examples. 

Our aim was to develop a neural network that would map a Fourier image ($\mathbf{x}$) of a black-hole silhouette with the visibility function given only in a subset of pixels to the angle ($\mathbf{y}$) between the black hole's spin and the line of sight of a distant observer. The description of the simulated dataset and the network's architecture is following.

\subsection{Simulated dataset}\label{dataset}

The dataset is obtained from real-space images which undergo a series of transformations resulting into a mock $uv$-plane with partial coverage.

\begin{figure}
   \centering
   \includegraphics[width=0.95\columnwidth]{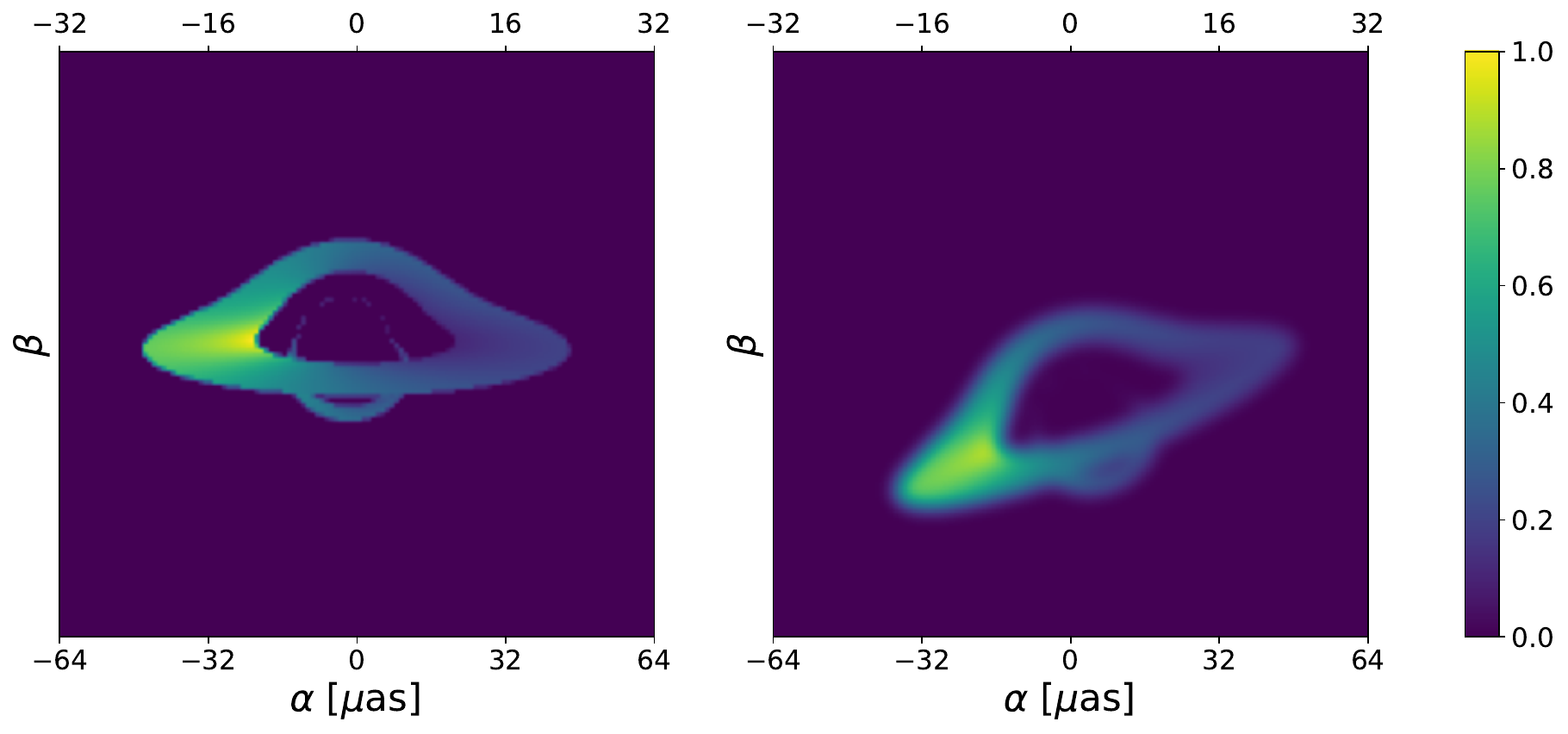}
   \caption{Silhouette of a Kerr black hole with rotation parameter $a=0.75$ against a counterrotating thin disk with outer radius $22.5GM/c^2$ seen at angle $\theta=78^\circ$ (\textit{left}) and the same silhouette distorted by a random translation, rotation, and a Gaussian blur (\textit{right}). Resolution of the images is $128\times 128$~pixels. The lower and upper $\alpha$-axes give two examples of physical scales: on the lower axis, $1~\mbox{pixel}=1~\mu\rm{as}$ while on the upper axis $1~\mbox{pixel}=0.5~\mu\rm{as}$. Corresponding scales in the $uv$-plane are given in Fig.~\ref{fourier}.}\label{augment}%
\end{figure}

\begin{figure}
   \centering
   \includegraphics[width=0.95\columnwidth]{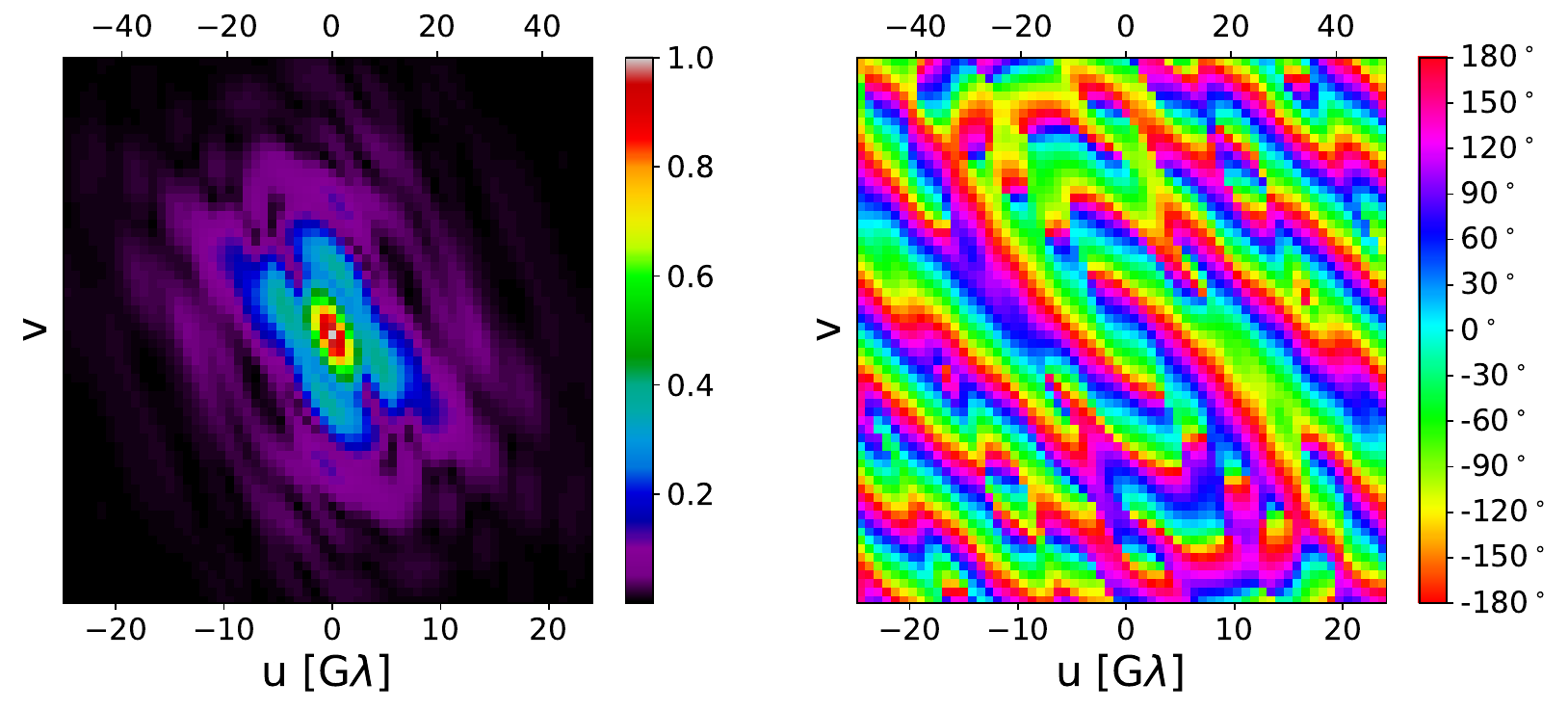}
   \caption{Amplitude (\textit{left}) and phase (\textit{right}) of the Fourier transform of the distorted image of Fig.~\ref{augment} (right panel). Resolution of the pictures is $64\times 64$~pixels. The upper and lower scales of the $u$-axis correspond to the physical scales of Fig.~\ref{augment} and are given in units of $10^9\lambda=1~\rm{G}\lambda$. On the lower axis, $1~\mbox{pixel}\approx0.8~\rm{G}\lambda$ while on the upper one, $1~\mbox{pixel}\approx1.6~\rm{G}\lambda$.}
              \label{fourier}%
\end{figure}

   \begin{figure*}
   \centering
   \includegraphics[width=\textwidth]{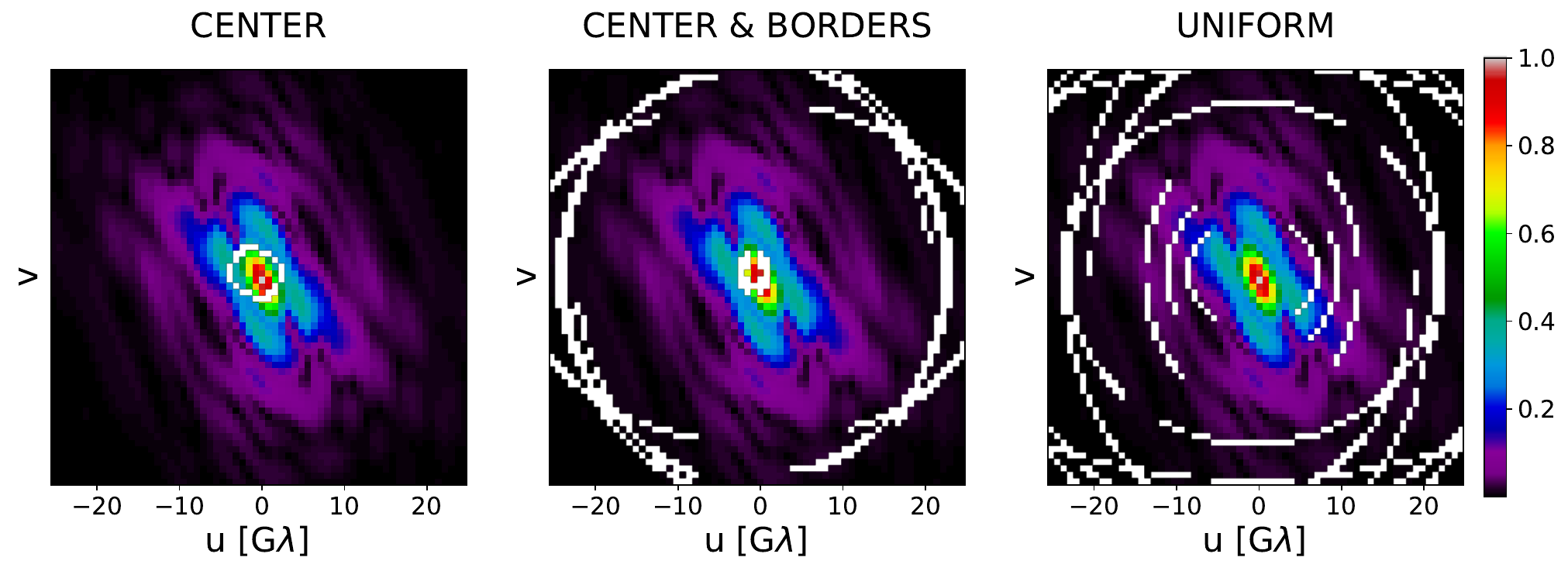}
   \caption{Amplitude of the Fourier transform with three types of masks overlaid for a low coverage. We require that the central part of the \texttt{center} and \texttt{center \& borders} patterns have the same coverage (here $\approx 24\%$ of the central part is covered). Absolute coverages of all patterns are $0.6\%$, $0.6\%$ ($10.3\%$ if the arcs are included), and $15\%$, respectively. The physical scale of the $u$-axis roughly corresponds to the scale of the EHT experiment. The \texttt{center} pattern mimics the coverage attained with EHT whereas cases \texttt{center \& borders} and \texttt{uniform} simulate the coverage of future interferometry experiments with additional dishes in higher and lower Earth orbits, respectively (see Subsec.~\ref{dataset} for details).}
              \label{coverage_015}%
    \end{figure*}

   \begin{figure*}
   \centering
   \includegraphics[width=\textwidth]{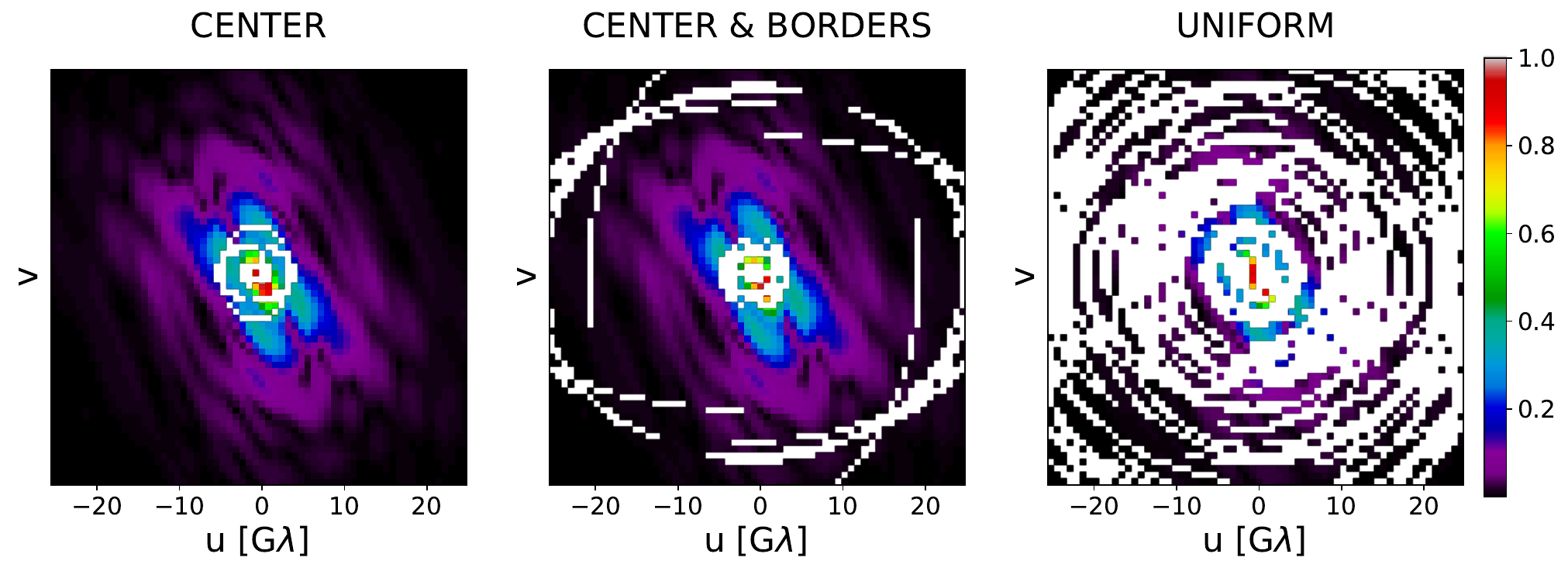}
   \caption{Amplitude of the Fourier transform with three types of masks overlaid for a low coverage. We require that the central part of the \texttt{center} and \texttt{center \& borders} patterns have the same coverage (here $\approx 75\%$ of the central part is covered). Absolute coverages of all patterns are $2\%$, $2\%$ ($12.7\%$ if the arcs are included), and $60\%$, respectively. Notation is that of Fig.~\ref{coverage_015}.}
 \label{coverage_060}%
    \end{figure*}

   \begin{figure}
   \centering
   \includegraphics[width=\columnwidth]{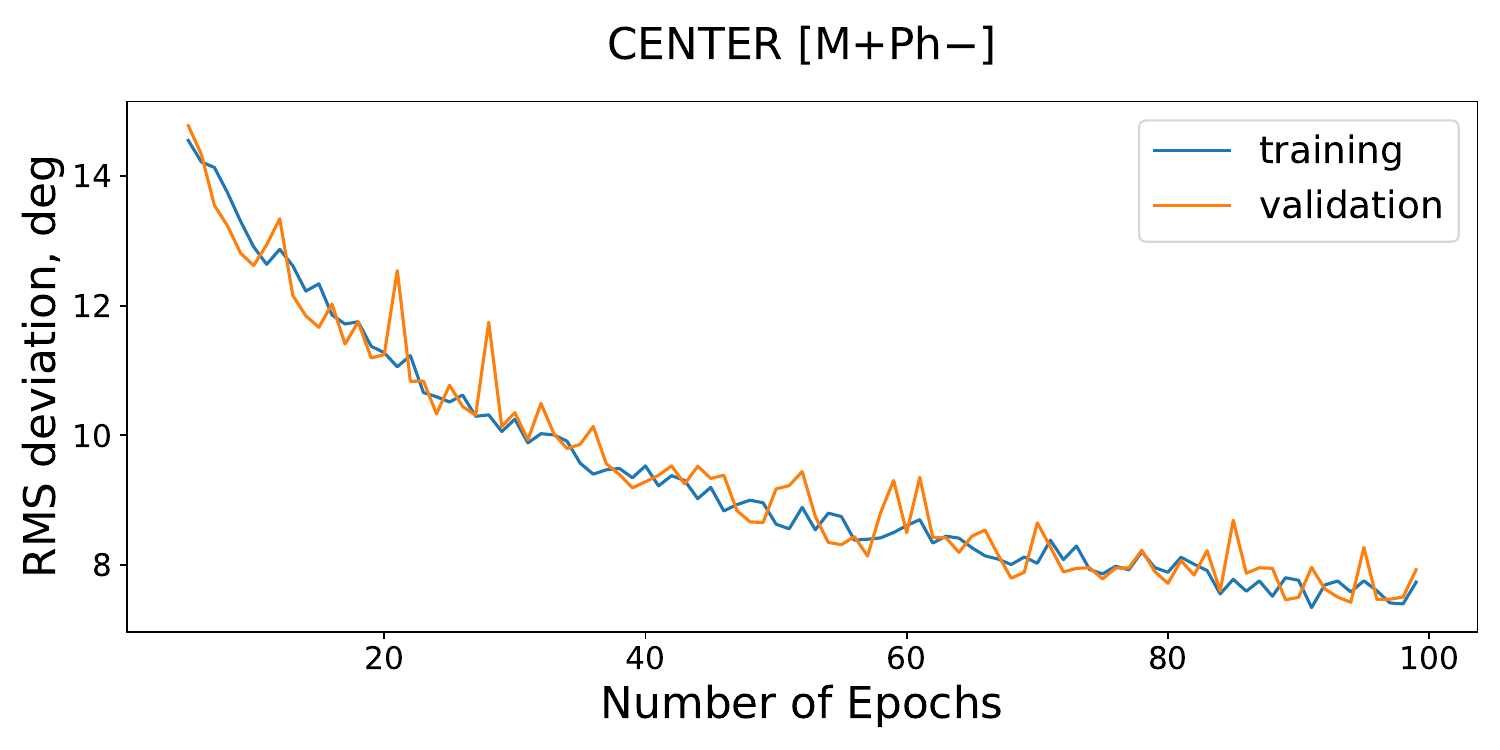}
   \caption{Learning curve for the case of \texttt{central} coverage and the mask as an extra input. Although dropout was set zero, there are no signs of overfitting.}
              \label{learning_curve}%
    \end{figure}

\begin{table}[htb!]
\caption{Parameter ranges for simulated pre-transformation images\label{par_ranges}}
\begin{tabular}{@{}*{3}{lcccc}}
\toprule
\multirow{3}{*}{} & inclination & outer  &  rotation  &   sense of   \\
 & angle,  & radius,  & parameter, & rotation \\
 & $\theta\,[{}^\circ]$ & $r_{\rm out}\,[M]$ & $a\,[M]$  & {} \\
\midrule
range & $[1,89]$ & $[15,24]$ & $[0,1]$ & +/- \\
increment & 1  & 2.25 & 0.25 &  random \\
\bottomrule
\end{tabular}
\end{table}

The real-space images are those of the silhouette of a Kerr black hole surrounded by a geometrically thin and optically thick accretion disk. We simulate the silhouette by tracing $256\times 256$ rays and then reduce the image to $128\times 128$ pixels by averaging over adjacent $2\times 2$ squares. We vary neither the distance to the black hole nor its mass, which fixes the angular scale of the image. In particular, the horizontal and vertical linear/angular scale $L$ is fixed as follows:
\begin{equation}
L = \frac{200}{3}\frac{GM}{c^2}\,, 
\end{equation}
where $M$ is the black hole's mass, $G$ is the gravitational constant, and $c$, the speed of light (hereafter, we set $G=c=1$). This results into the following relative linear and angular scales ($\Delta L$ and $\Delta\alpha$, respectively): 
\begin{equation}
\Delta L \approx 5\times10^{-3}\left(\frac{128}{N}\right)\left(\frac{M}{10^6M_\odot}\right)\,\frac{\mbox{AU}}{\mbox{pixel}}\,, 
\end{equation}
\begin{equation}
\Delta\alpha = 0.5\left(\frac{r_0}{10\,\mbox{kpc}}\right)\left(\frac{128}{N}\right)\left(\frac{M}{10^6M_\odot}\right)\,\frac{\mu\mbox{as}}{\mbox{pixel}}\,, 
\end{equation}
where $N$ is the 1D resolution of the real-space image and $r_0$, the distance to the black hole.

The elements of the disk are assumed to follow circular geodesic orbits with the inner radius of $10M$. The images are generated for a range of values of parameters which are the disk's outer radius~$r_{\rm out}$, Kerr rotation parameter~$a$, and the inclination angle~$\theta$. Also, the disk is chosen to be co- or counterrotating with probability $1/2$. The ranges of the parameters as well as their increments are given in Table~\ref{par_ranges}. There are 2,225 combinations of the parameters to which we add 76 images obtained from a trial simulation\,\footnote{Those are approximately evenly distributed among angles $1^\circ, 23^\circ, 45^\circ, 67^\circ, 89$ and generated for all the combinations of the outer radius and rotation parameter. They differ in the disk's sense of rotation.}. Thus, the total number of the images in training and validation sets before the transformations is 2,301. The training set batches are generated on the fly by applying random transformations (for more details on data augmentation, see below). A new batch is generated at each step of the training process. Each image in the batch is transformed randomly. The validation set batch is generated in the same way only once at the start of training and is used at every step. A test set that will be used to report the final results comprises 89 pre-transformation images with angles ranging from~$1^\circ$ to~$89^\circ$ in increments of~$1^\circ$ and the other parameters chosen randomly from their respective ranges (in a continuous manner), see Table~\ref{par_ranges}.

Then, we perform data augmentation on these images by applying translation, rotation and blur. The data augmentation is combined with a Fourier transform in the following order: rotation $\rightarrow$ Fouier transform $\rightarrow$ translation/blur. We carry out the translation and Gaussian blur in the $uv$-plane (see~\ref{gauss_fourier}) in order to avoid edge artifacts. This is especially convenient, because the input of the CNN is Fourier-transformed images.

The translations are by a (uniformly) random vector with the $x$- and $y$-components between $-15$ and $15$ pixels. The rotations are by an angle uniformly distributed between $-180^\circ$ and $180^\circ$. Finally, we apply a Gaussian blur (smoothing) with a sigma drawn from a uniform distribution, $\sigma\sim U(0,3\sqrt{3}M/\Delta L) = U(0,10.0)$. The maximum blur is chosen to be equal to the universal size of the shadow of a Schwarzschild black hole~\citet{Narayan_2019}. Fig.~\ref{augment} shows a silhouette of a Kerr black hole seen at $\theta=78^\circ$ and its version distorted by a translation, rotation, and a blur (to obtain the image, an inverse Fourier transform was applied after translating and blurring in the Fourier domain). Fig.~\ref{fourier} shows the original amplitude and phase of the image of Fig.~\ref{augment} (right panel). 

Note that, at the programming level, both real-space image and its Fourier transform are scale-free and their sizes are in pixels ($128\times 128$ for the real-space image and $64\times 64$ for the Fourier). The $uv$-image can be provided with a physical scale as follows:
\begin{equation}
\Delta\nu\,[\rm{G}\lambda/\rm{pixel}] = \frac{N-1}{k_{\rm pad}N^2}\frac{36\times 18}{\pi\,\Delta\alpha\,[\mu\rm{as}/pixel]}\,,
\label{ang_scale}
\end{equation}
where $\Delta\alpha$ and $N$ are, respectively, the physical scale and resolution of a real-space image, and $k_{\rm pad}$ is the zero-padding factor of the discrete Fourier transform used to make frequency bins narrower~\citep[e.g.][]{Donnelly_2005}. In this paper,  $k_{\rm pad}=2$, $N=128$. 

Fig.~\ref{augment} shows two examples of physical scales on the lower and upper axes: $1$ and~$0.5\,\mu{\rm as}/\rm{pixel}$, respectively. These correspondingly result in $\approx0.8$ and~$1.6\,\rm{G}\lambda/\rm{pixel}$ in Fig.~\ref{fourier}. The lower scale of Fig.~\ref{augment} is approximately equal to the scale of images obtained with EHT (cf. Fig.~3 in \citet{EHTI} and \citet{Akiyama_2015}).

Finally, the Fourier amplitude as well as the phase are overlaid with a mask that mimics the partial coverage of the $uv$-plane in observations. The mask is a $64\times 64$ matrix of ones and zeros, where the ones show which pixels are covered while the zeros encode the absence of observational signal. We use three types of masks which simulate different observational settings: a) only the central part of the Fourier image is covered, b) the coverage comprises the central part and a ring which is a few times bigger, and c) the coverage is more or less uniform over the $uv$-plane. We refer to these three cases as \texttt{center}, \texttt{center \& borders}, and \texttt{uniform}, respectively. Figs.~\ref{coverage_015} and~\ref{coverage_060} illustrate the three. 

\begin{figure}
   \begin{minipage}{\columnwidth}
   \centering
   \includegraphics[width=\columnwidth]{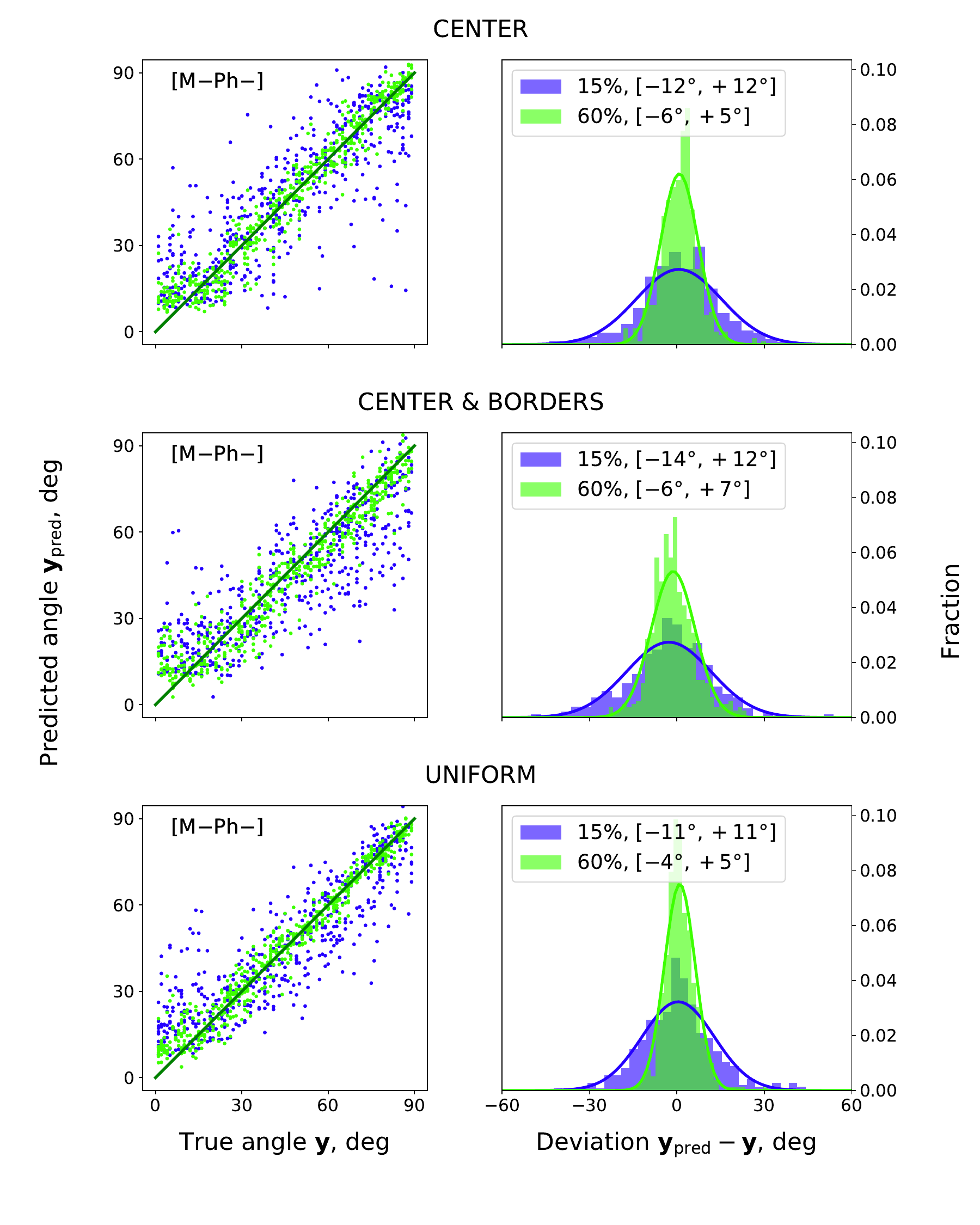}
   \caption{Angle as predicted by the M$-$Ph$-$ network vs. True angle (\textit{left column}) and distribution of errors for a low (\textit{bluish}) and high (\textit{green}) coverage (\textit{right column}). Solid curves surrounding the histograms are best-fit Gaussians. Standard deviations of the last are indicated in the legend. Three rows correspond to the three coverage patterns (see Figs.~\ref{coverage_015}, \ref{coverage_060} and Subsec.~\ref{dataset}).}
              \label{hist_mask=False_phase=False}%
	\end{minipage}
	   \begin{minipage}{\columnwidth}
   \centering
   \includegraphics[width=\columnwidth]{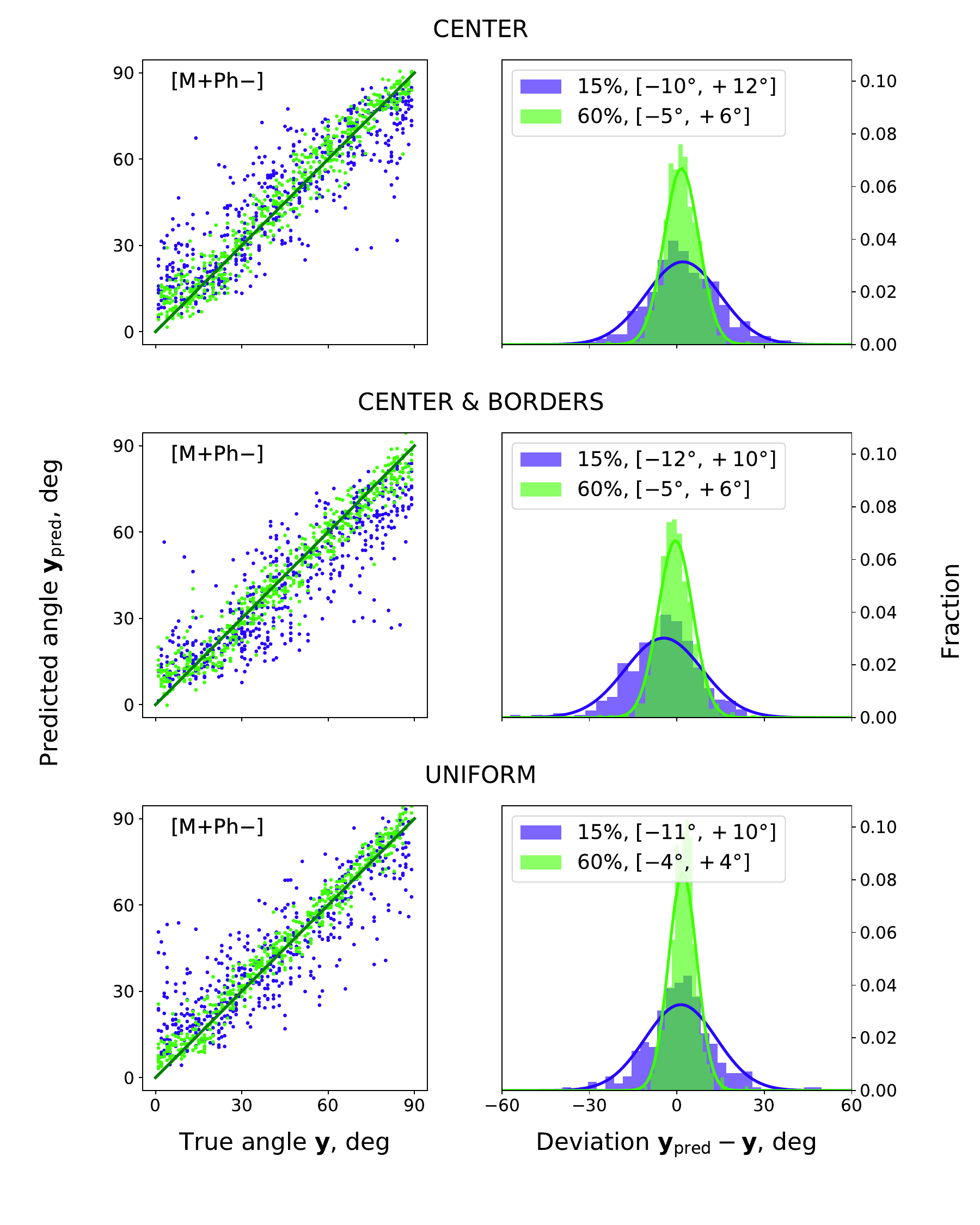}
   \caption{Angle as predicted by the M$+$Ph$-$ network vs. True angle. Notation is that of Fig.~\ref{hist_mask=False_phase=False}.}
              \label{hist_mask=True_phase=False}%
	\end{minipage}
    \end{figure}

\begin{figure}
   \begin{minipage}{\columnwidth}
   \centering
   \includegraphics[width=\columnwidth]{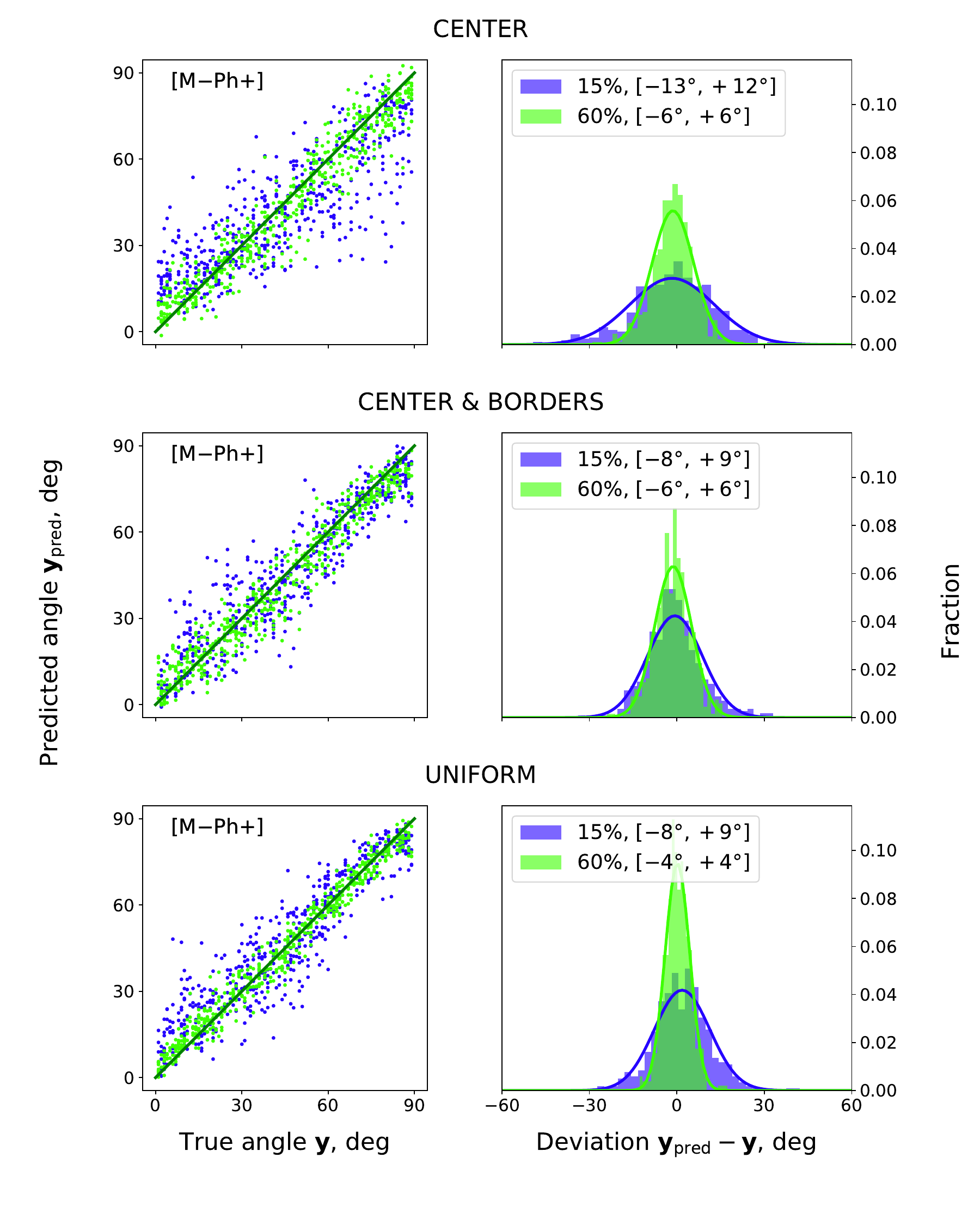}
   \caption{Angle as predicted by the M$-$Ph$+$ network vs. True angle. Notation is that of Fig.~\ref{hist_mask=False_phase=False}.}
              \label{hist_mask=False_phase=True}%
	\end{minipage}
	   \begin{minipage}{\columnwidth}
   \centering
   \includegraphics[width=\columnwidth]{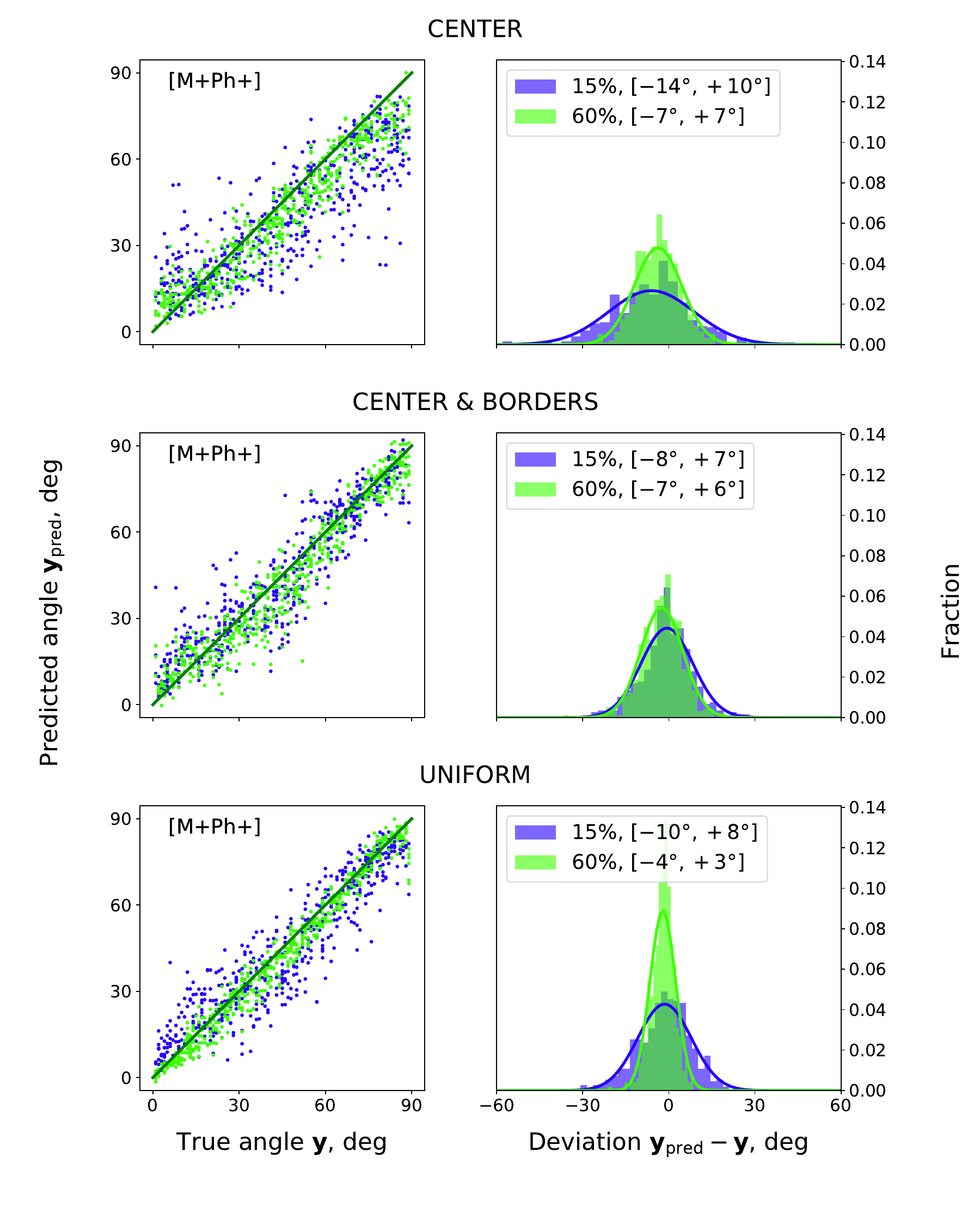}
   \caption{Angle as predicted by the M$+$Ph$+$ network vs. True angle. Notation is that of Fig.~\ref{hist_mask=False_phase=False}.}
              \label{hist_mask=True_phase=True}%
	\end{minipage}
    \end{figure}

\begin{figure*}[!htp]
\centering
   \includegraphics[width=\columnwidth]{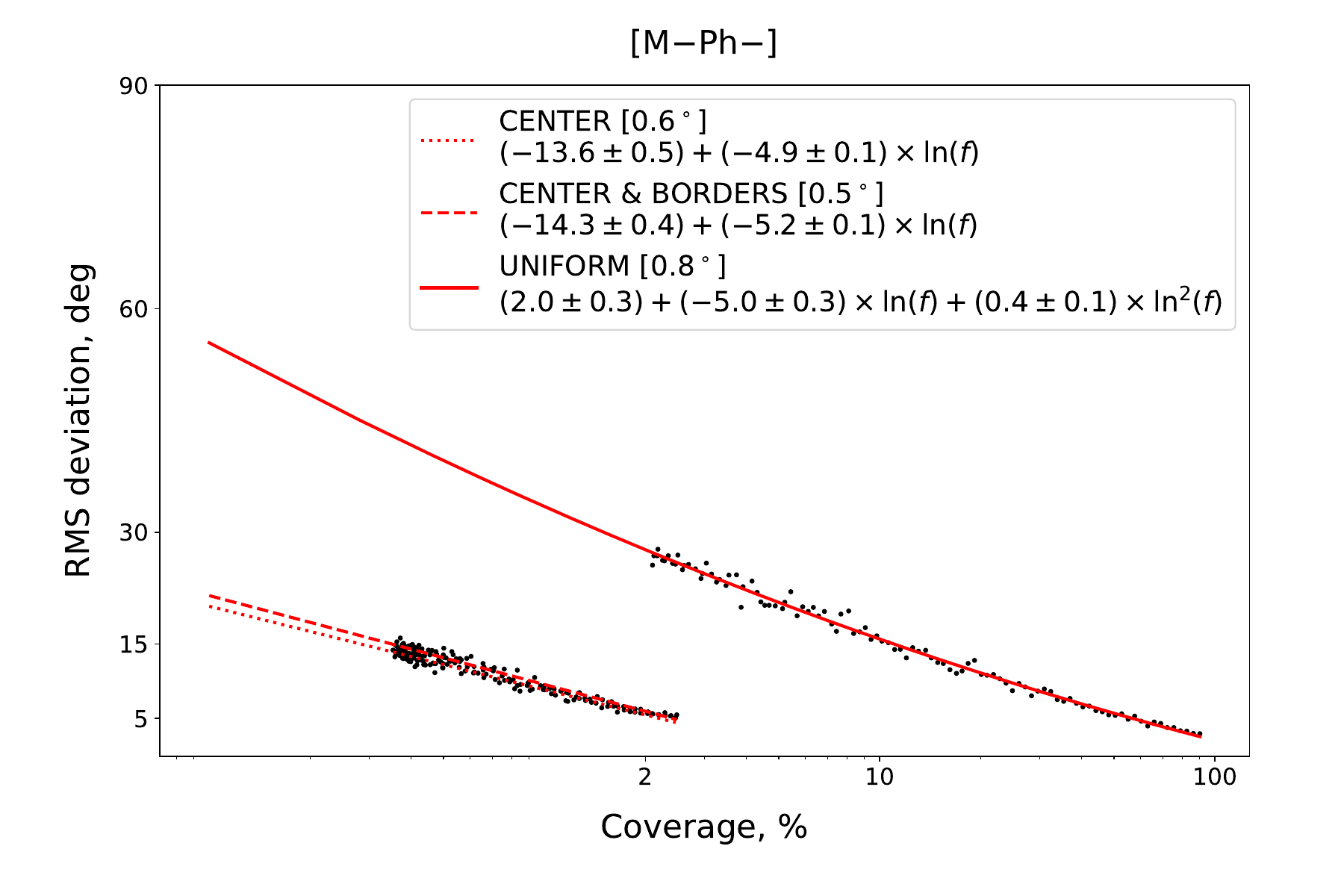}
   \includegraphics[width=\columnwidth]{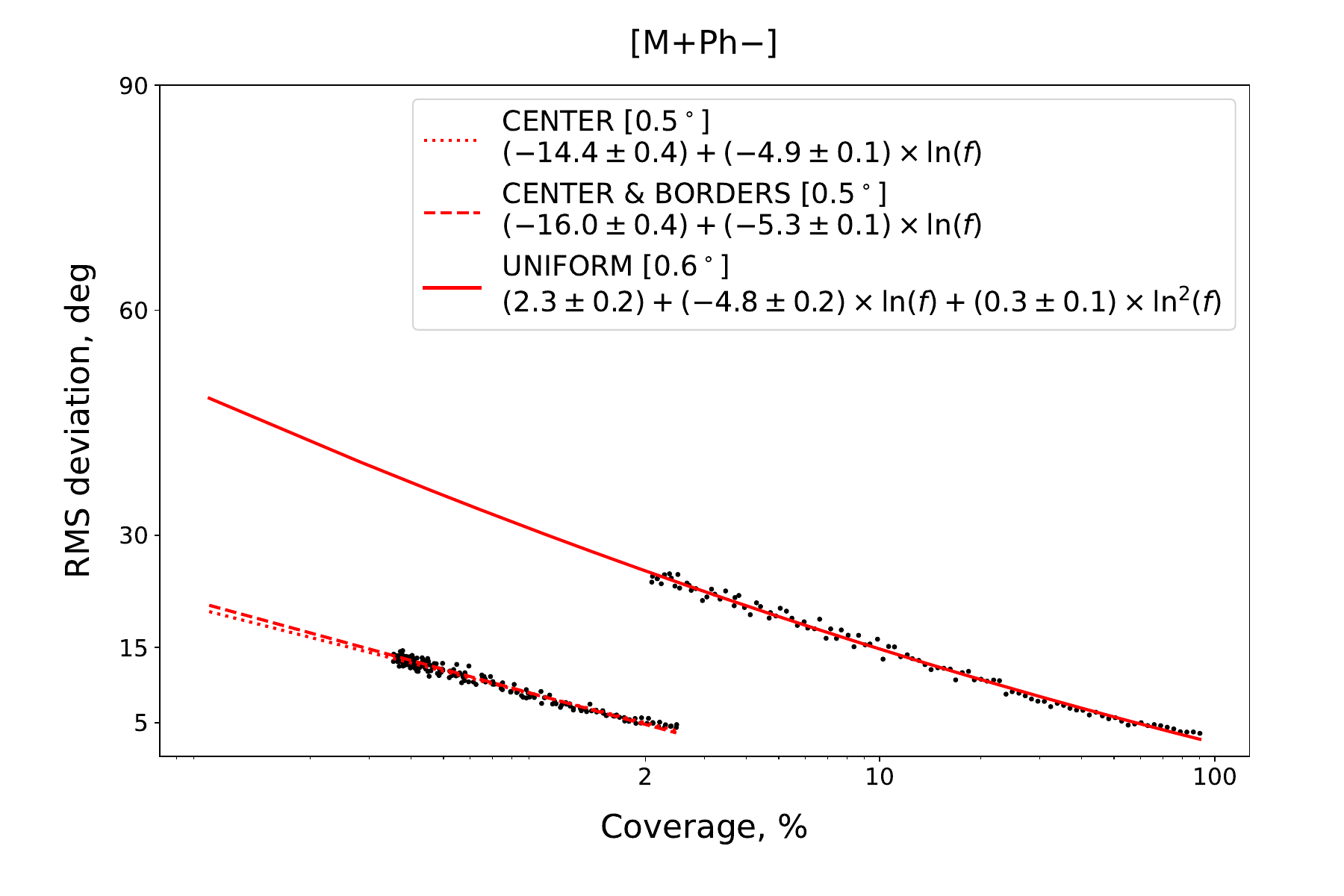}
   \includegraphics[width=\columnwidth]{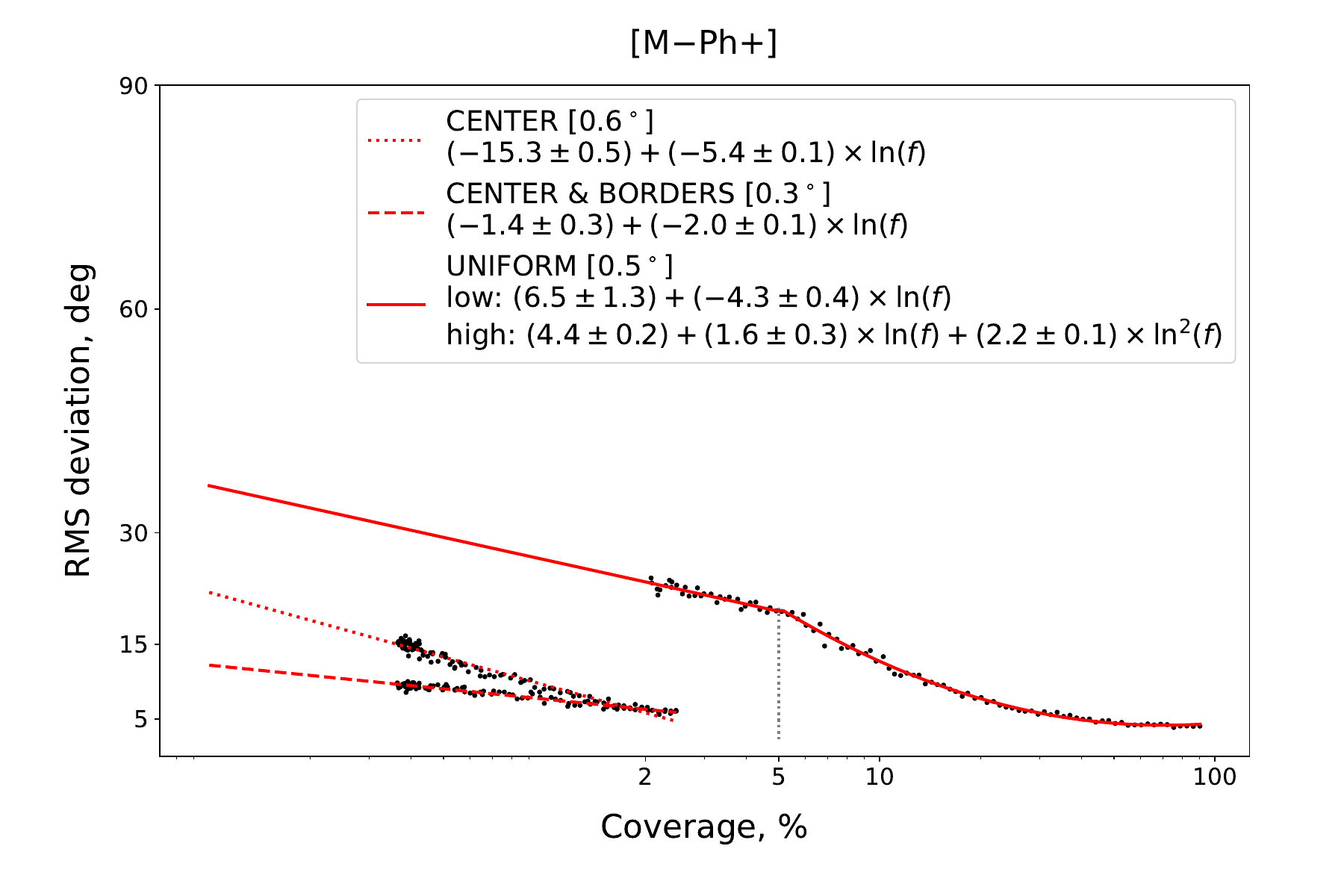}
   \includegraphics[width=\columnwidth]{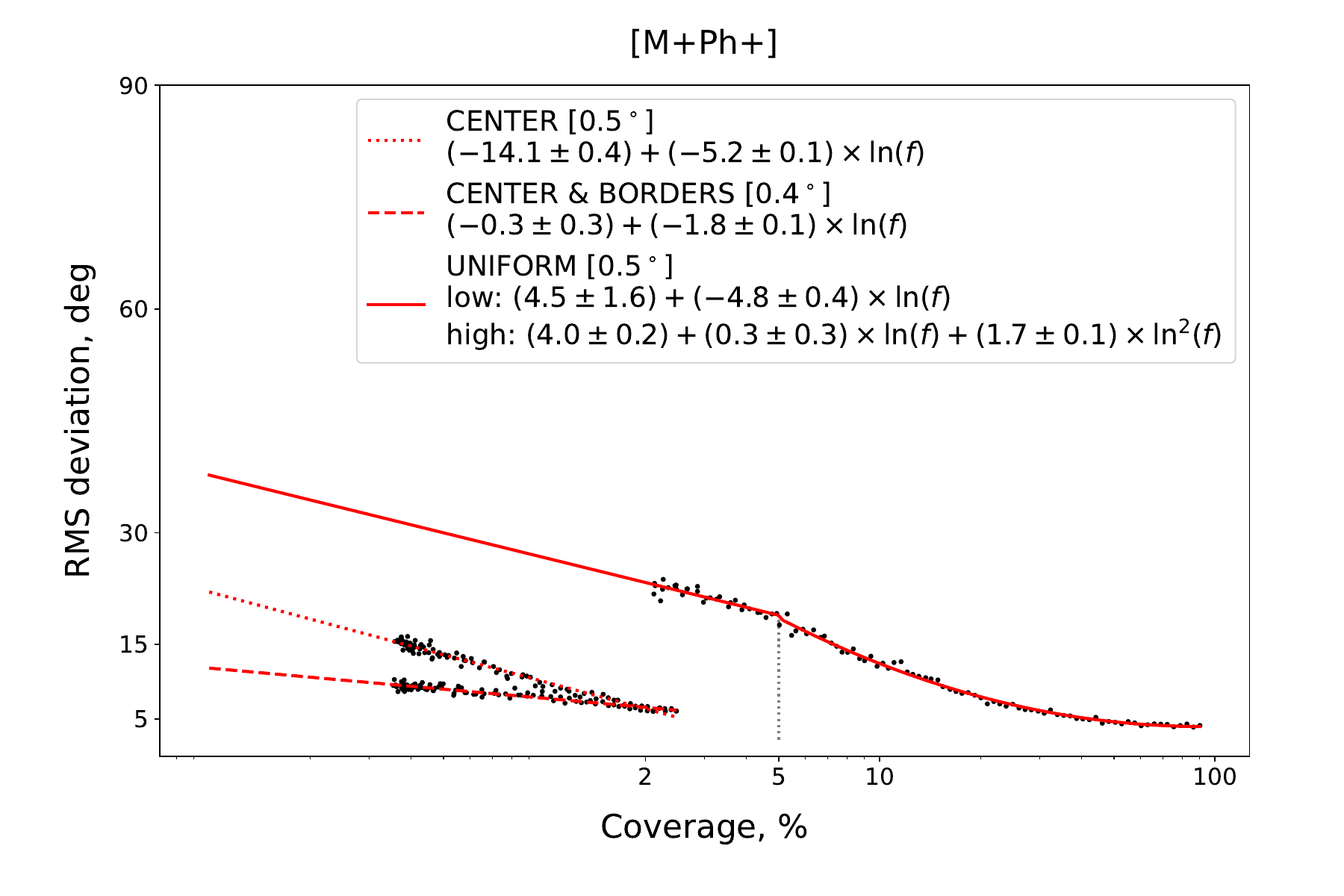}
   \caption{Semi-log plot of the RMS error as a function of the coverage for all four networks (see titles). Black dots show the error evaluated on a test sample of 512 images for each pattern and each coverage. Fits for the \texttt{center}, \texttt{center \& borders}, and \texttt{uniform} patterns are shown as a dotted, dashed, and solid line, respectively. The fitting linear or quadratic functions of $\ln{f}$ are given in the legend, with the fit's rms indicated in brackets. The error of networks that include the Fourier phase as an input is approximated with a linear function of $\ln{f}$ below $f=0.05$ and with a quadratic one, above.\label{cov_mask_all}}
\end{figure*}

To be able to compare the performance of the networks we describe in Subsect.~\ref{CNN_section}, we adopt the following convention for counting the coverage. For patterns \texttt{center} and \texttt{center \& borders} we require that the same fraction of the central part be covered and we refer to it as \textit{relative} coverage. The area of the part of the $uv$-plane we call central is, by definition, $40$ times smaller than the total area. On the other hand, the \textit{absolute} coverage for those patterns is the ratio of the number of pixels activated in the central area to the whole size of the $uv$-plane ($64\times 64$ pixels in this work). For the \texttt{center \& borders} pattern we choose not to include the pixels on the periphery, because we want to single out the effect of arcs when we will be evaluating the error of the networks. The relative coverage may be more illustrative in that it changes in a wider range while the absolute coverage of the \texttt{center} pattern cannot exceed $2.5\%$ (the maximum size of the central part). It is one more reason not to include the pixels on the periphery of the \texttt{center \& borders}, because there are about 10 times more of those pixels and their inclusion would lead to a very narrow interval on our final graph. In what follows we use the relative coverage in Figs.~\ref{hist_mask=False_phase=False}--%
\ref{hist_mask=True_phase=True} and the absolute coverage in final Fig.~\ref{cov_mask_all}.

Regarding the masks themselves, they consist of pairs of elliptic arcs symmetric w.r.t. the origin. In the \texttt{center} and \texttt{uniform} patterns as well as the central part of the \texttt{center \& borders} pattern, the radii\,\footnote{Hereafter, ``radius'' in relation to an elliptical arc means the geometric mean of its semi-major and semi-minor axes.} of the arcs are uniformly distributed between zero and a maximum value. The maximum value is $64\sqrt{2}/2$ for \texttt{uniform} and $\sqrt{1/40}\approx 0.16$ of that value for \texttt{center} and the central part of \texttt{center \& borders}. The angular sizes of the arcs are distributed normally with a mean of $2\pi/3$ and a standard deviation of $\pi/6$. The arcs' eccentricity follows the uniform distribution $U(0.1,0.9)$. For the \texttt{center \& borders} pattern, the radii of arcs on the periphery are distributed $\sim\mathcal{N}(32,0.64)$.

\begin{table}
      \caption{Architecture of the convolutional network. The number of parameters of the first layers differs in accordance with the numbers of channels in its input. If the mask is present in the input, it is fed through an extra channel. If the phase is present, its sine and cosine are fed through two extra channels. Presence/absence of mask and phase are denoted by M$\pm$ and Ph$\pm$, respectively.}
         \label{archi}
\begin{tabular}{@{}*{3}{llc}}
\toprule
                     Layer &        Output Shape & Number of Parameters \\
\midrule
\multirow{4}{*}{Conv2D} &  \multirow{4}{*}{(N, 62, 62, 16)} & 160 [M$-$Ph$-$]  \\
						& 			& 304 [M$+$Ph$-$] \\
						& 			& 448 [M$-$Ph$+$] \\
						& 			& 592 [M$+$Ph$+$] \\
MaxPooling2D &             (N, 31, 31, 16)        &              0        \\      
Conv2D &                          (N, 29, 29, 16)           &           2320     \\
MaxPooling2D      &       (N, 14, 14, 16)          &            0      \\        
Conv2D  &                         (N, 12, 12, 32)         &             4640      \\
MaxPooling2D  &             (N, 6, 6, 32)             &           0              \\
Flatten   &                         (N, 1152)           &                 0         \\     
Dense  &                            (N, 128)           &                  147584         \\
Dropout   &                        (N, 128)           &                  0              \\
Dense   &                            (N, 16)            &                  2064           \\
Dropout   &                        (N, 16)             &                 0            \\  
Dense   &                            (N, 1)          &                     17         \\    
\midrule
\multicolumn{2}{l}{\multirow{2}{*}{Activations:} }  &  ReLU (all but last)  \\
\multicolumn{2}{l}{\multirow{2}{*}{} } & LINEAR (last layer) \\
\midrule 
     \multicolumn{2}{l}{\multirow{4}{*}{Total (trainable) params:} }  &  156,785 [M$-$Ph$-$]    \\
     \multicolumn{2}{l}{\multirow{4}{*}{} } & 156,929 [M$+$Ph$-$] \\
     \multicolumn{2}{l}{\multirow{4}{*}{} } & 157,073 [M$-$Ph$+$] \\
     \multicolumn{2}{l}{\multirow{4}{*}{} } & 157,217 [M$+$Ph$+$] \\  
\bottomrule
\end{tabular}
\end{table}

If we adopt the physical scale of the lower $u$-axis of Fig.~\ref{fourier}, the \texttt{center} case roughly corresponds to the characteristic baselines of EHT (cf. Fig.~2 in \citet{EHTI}) while the \texttt{uniform} pattern mimics the prospective enhanced configurations of EHT with one or more small dishes in Low Earth Orbits (cf. Fig.~4 (third column) of \citet{Palumbo}). The \texttt{center \& borders} case is in turn characteristic of space-VLBI configurations that include dishes in higher orbits, e.g. in geosynchronous or medium Earth orbits \citep[][Fig.~1]{Fish_2019}, or in the Sun--Earth L2 point \citep{Kardashev_2014}. \textit{Note that these coverage masks and arcs therein are not identical to those simulated in the above-mentioned prospective space-VLBI experiments. The correspondence is rather qualitative.}

The dataset consisting of such masked Fourier images is then used to train a convolutional neural network.

\subsection{Convolutional neural network\label{CNN_section}}

A neural network that includes convolutional layers is known as convolutional neural network (CNN). In a convolutional layer, its input (an image) is ``scanned'' by many, typically, $3\times 3$ filters (kernels), with the output being the result of convolutions of the kernels with the respective parts of the image. As mentioned, such an architecture has proven to be extremely efficient in image recognition problems.

Table~\ref{archi} shows the full architecture of the CNN which we have developed. The code, the trained models' weights, and links to the training and test datasets are publicly available\,\footnote{\url{https://bitbucket.org/cosmoVlad/neuro-repo}}.

We compared four versions of this network which differed in their inputs. As one option, we turned on or off the mask, that is, either the mask was fed to the network as a separate input or not. These two cases are denoted by M$+$ and M$-$, respectively. In both cases the values of pixels that were out of coverage were set to zero. And the purpose of the mask as an extra input was an attempt to train the network to ignore the masked zero values and distinguish them from those that are part of the actual Fourier signal. For each of those cases, we also pass either only amplitudes of the Fourier images or phases as well. These cases are denoted by Ph$+$ and Ph$-$. To account for the periodicity of the phase, we passed its sine and cosine rather than the phase itself.

The four versions of the CNN were trained for $100$ epochs with batches comprising $64$ images and validated on a dataset of $2048$ images. As a loss function, we use the mean squared error (MSE),
\begin{equation}
\label{rms_eq}
\frac{1}{\mbox{batch size}}\sum\limits_{k=1}^{\mbox{\scriptsize{batch size}}}{\left(\theta_{\rm pred}^{(k)}- \theta^{(k)}\right)^2}
\end{equation}
Recall that each batch is generated on the fly by randomly choosing the respective number of pre-transformed images, applying transformations (rotation  $\rightarrow$ Fourier transform $\rightarrow$ translation/blur) with random parameters (except for Fourier transform), and overlaying a mask with a degree of relative coverage randomly and uniformly chosen between $0.1$ and $0.9$. A set of the random parameters is new each time an image is generated. Recall that, in the \texttt{center} and \texttt{center \& borders} cases, the relative coverage is the number of ones in the mask divided by the number of pixels in a central part of the image. The linear size of the central part is a free parameter, which was set to $\sqrt{1/40}\approx 0.16$ ($\approx 10$~pixels) in this work. The \texttt{center \& borders} case is different in that there are a few arcs added on the periphery of the image. The process of generating arcs is described in Subsect.~\ref{dataset}, and the number of arcs is also a free parameter, which was set to~$6$ in this work. The absolute coverage of a specific \texttt{center \& borders} pattern is obtained by dividing the number of activated pixels in the central part by the total number of pixels (that is, by $64\times 64$). In the \texttt{uniform} case the relative coverage coincides with the absolute one.

The degree of coverage of a single image was chosen randomly between $0.1$ and $0.9$ from a uniform distribution. The degree of coverage is defined as follows for different patterns. In the \texttt{uniform} case, it is the number of ones in the mask divided by the total number of pixels (that is, by $64\times 64$). In the \texttt{center} and \texttt{center \& borders} cases, it is the number of ones divided by the number of pixels in a central part of the image. The linear size of the central part is a free parameter, which was set to $0.16$ ($\approx 10$~pixels) in this work. The \texttt{center \& borders} case is different in that there are a few arcs added on the periphery of the image. 

Note that our CNN also contains dropout layers to prevent overfitting. However, we tried a few dropout rates between $0$ and $0.1$ and did not find any overfitting trend as dropout rate decreased. Fig.~\ref{learning_curve} shows a typical learning curve with zero dropout rate. These dropout layers may become useful when estimating confidence intervals with a technique described by~\citet{Levasseur_2017} (to be done elsewhere).

\section{Results and discussion}\label{results}

The efficiency of the four versions of the network is summarized in Figs.~\ref{hist_mask=False_phase=False}--\ref{hist_mask=True_phase=True} and Fig.~\ref{cov_mask_all}. Recall that these versions result from passing or not the mask and/or the phase as additional inputs to the network and are denoted as M$-$Ph$-$, M$+$Ph$-$, M$-$Ph$+$, M$+$Ph$+$.

The series of figures~\ref{hist_mask=False_phase=False}--\ref{hist_mask=True_phase=True} shows distributions of the discrepancy between a true angle and the answer given by the network. These distributions were evaluated on a dataset (test set) of 512 images, and each row represents the error evaluated on images of a different coverage pattern. The left panels are graphs Predicted angle vs. True angle, and the cumulative histograms of the deviations are on the right panels. The two colors represent the error distribution at two different degrees of relative coverage, $15\%$ and $60\%$. 

In the legend we indicate the $68\%$ quantile interval around the median of a historgram. For comparison we also draw best-fitting Gaussians, although the statistics of the deviations is not Gaussian as it becomes evident if one plots the $Q$--$Q$ plot or runs a normality test, e.g.~\citet{SHAPIRO_1965}. One reason not to expect the deviations to be Gaussian is that the angle cannot be negative by definition. This is manifested in how some two of the networks overestimate the angle at small values (note the elevated bottom left corner of the plot on the left panels of Figs.~\ref{hist_mask=False_phase=False} and~\ref{hist_mask=True_phase=False}). Also, in the \texttt{center} case all the versions of the network tend to underestimate angles that are close to the right angle. We defer the investigation of the statistical properties of such a network to future research.

We have also tested the performance of networks M$-$Ph$-$ and M$-$Ph$+$ on images with the maximal Gaussian blur (recall that the sigma of the blur is close to the universal size of the shadow of a Schwarzschild black hole and, thus, mimics a source nearing being unresolved; see also Sect.~\ref{dataset}). We have found that this significant blur does not affect the M$-$Ph$+$ network. In the case of M$-$Ph$-$, however, the network's performance worsens for all the three coverage patterns, with the last having the same error of about $\pm 15^\circ$ (at $60\%$ coverage). Such behavior is not unexpected, because the blur only affects the magnitude of the visibility function (see also \ref{gauss_fourier}). One interpretation of the same performance of M$-$Ph$+$ is that the network has learned to determine the angle mostly from the phase which is unaffected by smoothing. Meanwhile, the blur effectively cuts the large harmonics of the Fourier amplitude which makes covering anything other than the center of the Fourier image useless. This explains why the error of M$-$Ph$-$ becomes independent of the coverage pattern for large blurs.

Interestingly, the statistics of deviations of viewing angle in Deep Horizon \citep[][Fig.~5, top row]{deep_horizon} shows a similar pattern of overestimation at smaller angles and underestimation at larger ones, even though, in that paper, the very range of angles is restricted to $[15^\circ, 25^\circ]$. The effect is most apparent with larger Gaussian beam. If we adopt the physical scale of $\sim 1\;\mu\mbox{as}/\mbox{pixel}$ on Fig.~\ref{augment}, our maximum blur corresponds to a Gaussian beam of $\approx 10\mu\mbox{as}$. Since we evaluate the error on a dataset with randomly generated individual blurs, we can take a blur of $\sim 5\mu\mbox{as}$ as an estimate of average blur in the dataset. Comparing the respective columns of \citep[][Fig.~5, top row]{deep_horizon} we see that the error of our best M$-$Ph$+$ network is about twice as high at the higher coverage and with the smaller Gaussian beam and is $1--1.5$ times as highwith the larger Gaussian beam. This network of ours, however, does not suffer from overestimation at these degrees of coverage and comprises the entire range of angles rather than $[15^\circ, 25^\circ]$. 

Our network is also more universal in that it is trained on a set of degrees of coverage. It is hard to compare it directly to Deep Horizon, because the latter was trained directly on real-space images, which does not seem to take into account the deconvolution process from a $uv$-image with particular coverage. In addition, the M$-$Ph$+$ network is basically not sensitive to blur as we explained above.

Figs.~\ref{hist_mask=False_phase=False}--\ref{hist_mask=True_phase=True} and Fig.~\ref{cov_mask_all} lead to the following conclusions. First, as the degree of coverage for a given pattern increases, the standard error decreases, which one might naturally expect. Second, in terms of \textit{relative} coverage the \texttt{uniform} pattern leads to a lower error than the other two patterns. However, in terms of \textit{absolute} coverage, the error of \texttt{center} and \texttt{center \& borders} is definitely smaller than that of \texttt{uniform}. At the same time, although the \texttt{central} patterns outperform the \texttt{uniform} pattern at low degrees of absolute coverage, they suffer from overestimation at lower angles. The uniform pattern cures this problem in almost all the cases considered (except for the M$-$Ph$-$ network). Third, the introduction of input phase alone improves the quality of the networks on the \texttt{center \& borders} and \texttt{uniform} patterns. It reduces the error and improves the statistical quality of error distribution making less skewed (no underestimation at lower angles). The introduction of input mask alone appears to be beneficial, too, but mostly for the statistical quality (the skewness is reduced). Somewhat surprisingly, the network combining phase and mask performs worse than M$-$Ph$+$. All in all, the M$-$Ph$+$ demonstrate the best results from the point of view of both error magnitude and statistical quality.

Fig.~\ref{cov_mask_all} shows how the standard error depends on the degree of absolute coverage for all the three patterns. The black dots show the errors evaluated on a dataset (test set) of 512 images while the lines (dotted, dashed, and solid) are graphs of fitting polynomials (note that the plot is semi-log). The polynomials are linear for \texttt{center} and \texttt{center \& borders} and either quadratic or piecewise linear-quadratic for \texttt{uniform}.

These graphs illustrate a few tendencies. First, again, the standard error decreases as the degree of coverage increases. Second, at the same level of (small) absolute coverage, the \texttt{central} cases demonstrate better results than the \texttt{uniform} one. Third, \texttt{center} with and without \texttt{borders} produce approximately the same error if there is no input phase. Otherwise, for the Ph$+$ networks,the addition of the borders decreases the error approximately twice at the lowest coverages. At the maximum degree of absolute coverage for these cases ($2.5\%$), the errors are the same. Finally, the dependence Error vs. Coverage shows an interesting feature for the networks with input phase and \texttt{uniform} pattern -- starting from a coverage of $5\%$ the error drops faster than at degrees of coverage $<5\%$. For that reason, we choose a piecewise polynomial function to fit the results in those cases.

These results indicate that, if the inclination angle is not too small, in order to determine it with these networks within, for example, $10^\circ$, the use of Earth-sized baselines is sufficient. For the same level of (low) absolute coverage, the error can be improved by adding long space baselines (this would also add more phase closure conditions and, thus, more information on Fourier phase). For small angles (face-on orientation of the disk), space configuration with preferably uniform coverage should be used. Using the polynomial fits, we find typical values of $0.8\%$ (\texttt{center} and \texttt{center \& borders}) and $25\%$ (\texttt{uniform}) leading to an error of $10^\circ$ for the networks without input phase. For the Ph$+$ networks, these values are $1\%$, $0.3\%$, and $14\%$.

The fact that the angle can be determined from probing the central part of the $uv$-plane alone may be explained by the presence of a characteristic scale associated with each angle. If this is the case, one only needs to probe the first zero of the visibility function. Then, covering the central part is sufficient, provided that the inclination angle scale resides in it. The classical example here is the measurement of Betelgeuse's diameter by~\citet{Michelson_1921}. A black hole shadow can also be approximated by a 4-parametric crescent model which suggests a characteristic scale in the visibility function~\citep{Kamruddin_2013}. Such a scale may also explain why the addition of the phase does not significantly improve the performance of the networks on images with low blur. This is because the networks learn the scale already from the magnitude of the visibility function, and, since there are supposedly no other scales associated with the inclination angle, the same scale is present in the phase, which does not bring anything new. The phase could be useful, though, if the blur is significant as we saw above. Recall that the blur introduces a cut-off in the magnitude but leaves the phase intact. In this case the networks with input phase significantly outperfrom the ones without.

To summarize, we developed a proof-of-concept convolutional neural network which infers the angle of inclination of a Kerr black hole from the partially covered $uv$-plane of the shadow of the black hole against a geometrically thin and optically thick accretion disk. We explicitly found how the network's error depends on the degree of coverage of the $uv$-plane and compared four different versions of the network on three different types of input. We showed that the best results in terms of both error and statistics are attained by the network which takes both amplitude and phase as inputs and operates on Fourier images with uniform coverage.

Although the performance of this proof-of-concept network indicates that the enhanced configurations of EHT with dishes in Low Earth Orbits might be a better choice of future observations, this result needs to be elaborated on to be applicable to real observations. In future research we plan to develop a more sophisticated version of the network to be trained on more realisic images, such as resulting from one of the codes in~\citep{harms}. Also, a separate work is required to study the statistical properties of the deviations, given that the over- and underestimation patterns we saw for some versions of the network are similar to those in Deep Horizon.

\section{Acknowledgements}

For funding information, see the journal version of this paper~\citep{2021A&C....3600467P}.

We are grateful to V.S. Beskin and Yu.Yu. Kovalev for providing a work en\-viron\-ment which made completion of this paper possible. One of us (VNS) thanks A.~Alakoz for educational remarks on radio interferometry and A.~Radkevich, V.~Kozin, and S.~Repin for providing conditions favorable to research.

We also made use of Keras, \citep{keras_repo}, Theano \citep{theano}, TensorFlow \citep{tensorflow}, IPython \citep{PER-GRA:2007}, SciPy \citep{jones_scipy_2001},  Matplotlib \citep{Hunter:2007}, NumPy \citep{van2011numpy}.

\appendix

\section{Translation and Gaussian blur in the Fourier domain\label{gauss_fourier}}

In order to translate a real-space image by $\Delta x$ and $\Delta y$ in the horizontal and/or vertical directions, respectively, one should add a correction to its Fourier phase~\citep[e.g.][]{recipes},
\begin{eqnarray}
\Delta\arg{\mathcal{V}(i,j)} &=& \frac{2\pi}{k_{\rm pad}N}\left[-u_i\Delta x + v_j\Delta y\right]\,, \\
u_i&\equiv& i - \frac{k_{\rm pad}N}{2}\,, \\
v_j&\equiv& j - \frac{k_{\rm pad}N}{2}\,,
\end{eqnarray}
where $i,j = 0,1,2,\ldots,(k_{\rm pad}N-1)$, $k_{\rm pad}$ is the padding factor, and $N$, the number of pixels along each dimension of the real-space image (see also notation following eq.~(\ref{ang_scale})). The choice of signs in the expression for phase is consistent with the details of the numerical implementation of the Fast Fourier Transform algorithm~\citep{recipes,van2011numpy} and our assumptions that $\Delta x>0$ and $\Delta y>0$ imply translation to the right and upward. 

In the real space, Gaussian blur is a convolution of the Gaussian kernel with the image. By the convolution theorem~\citep[e.g.][]{recipes}, it is pixel-wise multiplication in the Fourier domain:
\begin{eqnarray}
|\mathcal{V}_{\rm blur}(i,j)| &=& |\mathcal{V}(i,j)| \nonumber \\
{}&\times& \exp{\left[-2\left(\frac{\pi\sigma}{k_{\rm pad}N}\right)^2\left(u_i^2 + v_j^2\right)\right]}\,, \nonumber \\
\end{eqnarray}
where $\sigma$ is the standard deviation of the Gaussian kernel in pixel units.

\bibliography{refs_new}

\begin{thebibliography}{66}
\expandafter\ifx\csname natexlab\endcsname\relax\def\natexlab#1{#1}\fi
\providecommand{\url}[1]{\texttt{#1}}
\providecommand{\href}[2]{#2}
\providecommand{\path}[1]{#1}
\providecommand{\DOIprefix}{doi:}
\providecommand{\ArXivprefix}{arXiv:}
\providecommand{\URLprefix}{URL: }
\providecommand{\Pubmedprefix}{pmid:}
\providecommand{\doi}[1]{\href{http://dx.doi.org/#1}{\path{#1}}}
\providecommand{\Pubmed}[1]{\href{pmid:#1}{\path{#1}}}
\providecommand{\bibinfo}[2]{#2}
\ifx\xfnm\relax \def\xfnm[#1]{\unskip,\space#1}\fi
\bibitem[{Abadi et~al.(2015)Abadi, Agarwal, Barham, Brevdo, Chen, Citro,
  Corrado, Davis, Dean, Devin, Ghemawat, Goodfellow, Harp, Irving, Isard, Jia,
  Jozefowicz, Kaiser, Kudlur, Levenberg, Man\'{e}, Monga, Moore, Murray, Olah,
  Schuster, Shlens, Steiner, Sutskever, Talwar, Tucker, Vanhoucke, Vasudevan,
  Vi\'{e}gas, Vinyals, Warden, Wattenberg, Wicke, Yu and Zheng}]{tensorflow}
\bibinfo{author}{Abadi, M.}, \bibinfo{author}{Agarwal, A.},
  \bibinfo{author}{Barham, P.}, et~al., \bibinfo{year}{2015}.
\newblock \bibinfo{title}{{TensorFlow}: Large-scale machine learning on
  heterogeneous systems}.
\newblock \URLprefix \url{https://www.tensorflow.org/}. \bibinfo{note}{software
  available from tensorflow.org}.
\bibitem[{Agarwal et~al.(2018)Agarwal, Dav{\'{e}} and Bassett}]{Agarwal_2018}
\bibinfo{author}{Agarwal, S.}, \bibinfo{author}{Dav{\'{e}}, R.},
  \bibinfo{author}{Bassett, B.A.}, \bibinfo{year}{2018}.
\newblock \bibinfo{title}{Painting galaxies into dark matter haloes using
  machine learning}.
\newblock \bibinfo{journal}{Monthly Notices of the Royal Astronomical Society}
  \bibinfo{volume}{478}, \bibinfo{pages}{3410--3422}.
\newblock \DOIprefix\doi{10.1093/mnras/sty1169}.
\bibitem[{Akiyama et~al.(2015)Akiyama, Lu, Fish, Doeleman, Broderick, Dexter,
  Hada, Kino, Nagai, Honma, Johnson, Algaba, Asada, Brinkerink, Blundell,
  Bower, Cappallo, Crew, Dexter, Dzib, Freund, Friberg, Gurwell, Ho, Inoue,
  Krichbaum, Loinard, MacMahon, Marrone, Moran, Nakamura, Nagar, Ortiz-Leon,
  Plambeck, Pradel, Primiani, Rogers, Roy, SooHoo, Tavares, Tilanus, Titus,
  Wagner, Weintroub, Yamaguchi, Young, Zensus and Ziurys}]{Akiyama_2015}
\bibinfo{author}{Akiyama, K.}, \bibinfo{author}{Lu, R.S.},
  \bibinfo{author}{Fish, V.L.}, et~al., \bibinfo{year}{2015}.
\newblock \bibinfo{title}{230 ghz {VLBI} observations of {M87}:
  event-horizon-scale structure during an enhanced very-high-energy
  $\gamma$-ray state in 2012}.
\newblock \bibinfo{journal}{The Astrophysical Journal} \bibinfo{volume}{807},
  \bibinfo{pages}{150}.
\newblock \DOIprefix\doi{10.1088/0004-637x/807/2/150}.
\bibitem[{Alibert and Venturini(2019)}]{Alibert_2019}
\bibinfo{author}{Alibert, Y.}, \bibinfo{author}{Venturini, J.},
  \bibinfo{year}{2019}.
\newblock \bibinfo{title}{Using deep neural networks to compute the mass of
  forming planets}.
\newblock \bibinfo{journal}{Astronomy {\&} Astrophysics} \bibinfo{volume}{626},
  \bibinfo{pages}{A21}.
\newblock \DOIprefix\doi{10.1051/0004-6361/201834942}.
\bibitem[{{An} et~al.(2019){An}, {Hong}, {Zheng}, {Ye}, {Qian}, {Fu}, {Guo},
  {Jaiswal}, {Kong}, {Lao}, {Liu}, {Liu}, {Lv}, {Mohan}, {Shen}, {Wang}, {Wu},
  {Wu}, {Zhang}, {Zhang}, {Zheng} and {Zhong}}]{Hong_2019}
\bibinfo{author}{{An}, T.}, \bibinfo{author}{{Hong}, X.},
  \bibinfo{author}{{Zheng}, W.}, et~al., \bibinfo{year}{2019}.
\newblock \bibinfo{title}{{Progress and perspective of space VLBI in China}}.
\newblock \bibinfo{journal}{arXiv e-prints}
  \href{http://arxiv.org/abs/1901.07796}{\tt arXiv:1901.07796}.
\bibitem[{Askar et~al.(2019)Askar, Askar, Pasquato and Giersz}]{Askar_2019}
\bibinfo{author}{Askar, A.}, \bibinfo{author}{Askar, A.},
  \bibinfo{author}{Pasquato, M.}, et~al., \bibinfo{year}{2019}.
\newblock \bibinfo{title}{Finding black holes with black boxes {\textendash}
  using machine learning to identify globular clusters with black hole
  subsystems}.
\newblock \bibinfo{journal}{Monthly Notices of the Royal Astronomical Society}
  \bibinfo{volume}{485}, \bibinfo{pages}{5345--5362}.
\newblock \DOIprefix\doi{10.1093/mnras/stz628}.
\bibitem[{Berger and Stein(2018)}]{Berger_2018}
\bibinfo{author}{Berger, P.}, \bibinfo{author}{Stein, G.},
  \bibinfo{year}{2018}.
\newblock \bibinfo{title}{A volumetric deep convolutional neural network for
  simulation of mock dark matter halo catalogues}.
\newblock \bibinfo{journal}{Monthly Notices of the Royal Astronomical Society}
  \bibinfo{volume}{482}, \bibinfo{pages}{2861--2871}.
\newblock \DOIprefix\doi{10.1093/mnras/sty2949}.
\bibitem[{Bishop(2006)}]{bishop}
\bibinfo{author}{Bishop, C.M.}, \bibinfo{year}{2006}.
\newblock \bibinfo{title}{Pattern Recognition and Machine Learning (Information
  Science and Statistics)}.
\newblock \bibinfo{publisher}{Springer-Verlag}, \bibinfo{address}{Berlin,
  Heidelberg}.
\bibitem[{Brinkerink et~al.(2016)Brinkerink, Müller, Falcke, Bower, Krichbaum,
  Castillo, Deller, Doeleman, Fraga-Encinas, Goddi,
  Hern{\'{a}}ndez-G{\'{o}}mez, Hughes, Kramer, L{\'{e}}on-Tavares, Loinard,
  Monta{\~{n}}a, Mo{\'{s}}cibrodzka, Ortiz-Le{\'{o}}n, Sanchez-Arguelles,
  Tilanus, Wilson and Zensus}]{Brinkerink_2016}
\bibinfo{author}{Brinkerink, C.D.}, \bibinfo{author}{Müller, C.},
  \bibinfo{author}{Falcke, H.}, et~al., \bibinfo{year}{2016}.
\newblock \bibinfo{title}{Asymmetric structure in sgr~a* at 3~mm from closure
  phase measurements with {VLBA}, {GBT} and {LMT}}.
\newblock \bibinfo{journal}{Monthly Notices of the Royal Astronomical Society}
  \bibinfo{volume}{462}, \bibinfo{pages}{1382--1392}.
\newblock \DOIprefix\doi{10.1093/mnras/stw1743}.
\bibitem[{{Broderick} and {Event Horizon Telescope
  Collaboration}(2020)}]{MCMC_Broderick}
\bibinfo{author}{{Broderick}, A.E.}, \bibinfo{author}{{Event Horizon Telescope
  Collaboration}}, \bibinfo{year}{2020}.
\newblock \bibinfo{title}{{THEMIS: A Parameter Estimation Framework for the
  Event Horizon Telescope}}.
\newblock \bibinfo{journal}{\apj} \bibinfo{volume}{897}, \bibinfo{pages}{139}.
\newblock \DOIprefix\doi{10.3847/1538-4357/ab91a4}.
\bibitem[{Chollet(2015)}]{keras_repo}
\bibinfo{author}{Chollet, F.}, \bibinfo{year}{2015}.
\newblock \bibinfo{title}{{Keras}}.
\newblock \URLprefix \url{https://github.com/fchollet/keras}.
\bibitem[{{Chollet}(2017)}]{keras}
\bibinfo{author}{{Chollet}, F.}, \bibinfo{year}{2017}.
\newblock \bibinfo{title}{{Deep learning with Python}}.
\newblock \bibinfo{edition}{1st} ed., \bibinfo{publisher}{Manning
  Publications}.
\bibitem[{{Clark}(1980)}]{CLEAN-2}
\bibinfo{author}{{Clark}, B.G.}, \bibinfo{year}{1980}.
\newblock \bibinfo{title}{{An efficient implementation of the algorithm
  'CLEAN'}}.
\newblock \bibinfo{journal}{\aap} \bibinfo{volume}{89}, \bibinfo{pages}{377}.
\bibitem[{Clery(2012)}]{EHT-2}
\bibinfo{author}{Clery, D.}, \bibinfo{year}{2012}.
\newblock \bibinfo{title}{Worldwide telescope aims to look into milky way
  galaxy's black heart}.
\newblock \bibinfo{journal}{Science} \bibinfo{volume}{335},
  \bibinfo{pages}{391--391}.
\newblock \DOIprefix\doi{10.1126/science.335.6067.391}.
\bibitem[{Cybenko(1989)}]{Cybenko_1989}
\bibinfo{author}{Cybenko, G.}, \bibinfo{year}{1989}.
\newblock \bibinfo{title}{Approximation by superpositions of a sigmoidal
  function}.
\newblock \bibinfo{journal}{Mathematics of Control, Signals, and Systems}
  \bibinfo{volume}{2}, \bibinfo{pages}{303--314}.
\newblock \DOIprefix\doi{10.1007/bf02551274}.
\bibitem[{Doeleman et~al.(2012)Doeleman, Fish, Schenck, Beaudoin, Blundell,
  Bower, Broderick, Chamberlin, Freund, Friberg, Gurwell, Ho, Honma, Inoue,
  Krichbaum, Lamb, Loeb, Lonsdale, Marrone, Moran, Oyama, Plambeck, Primiani,
  Rogers, Smythe, SooHoo, Strittmatter, Tilanus, Titus, Weintroub, Wright,
  Young and Ziurys}]{Doeleman_2012}
\bibinfo{author}{Doeleman, S.S.}, \bibinfo{author}{Fish, V.L.},
  \bibinfo{author}{Schenck, D.E.}, et~al., \bibinfo{year}{2012}.
\newblock \bibinfo{title}{Jet-launching structure resolved near the
  supermassive black hole in m87}.
\newblock \bibinfo{journal}{Science} \bibinfo{volume}{338},
  \bibinfo{pages}{355--358}.
\newblock \DOIprefix\doi{10.1126/science.1224768}.
\bibitem[{Doeleman et~al.(2008)Doeleman, Weintroub, Rogers, Plambeck, Freund,
  Tilanus, Friberg, Ziurys, Moran, Corey, Young, Smythe, Titus, Marrone,
  Cappallo, Bock, Bower, Chamberlin, Davis, Krichbaum, Lamb, Maness, Niell,
  Roy, Strittmatter, Werthimer, Whitney and Woody}]{Doeleman_SgrA}
\bibinfo{author}{Doeleman, S.S.}, \bibinfo{author}{Weintroub, J.},
  \bibinfo{author}{Rogers, A.E.E.}, et~al., \bibinfo{year}{2008}.
\newblock \bibinfo{title}{Event-horizon-scale structure in the supermassive
  black hole candidate at the galactic centre}.
\newblock \bibinfo{journal}{Nature} \bibinfo{volume}{455},
  \bibinfo{pages}{78--80}.
\newblock \DOIprefix\doi{10.1038/nature07245}.
\bibitem[{Donnelly and Rust(2005)}]{Donnelly_2005}
\bibinfo{author}{Donnelly, D.}, \bibinfo{author}{Rust, B.},
  \bibinfo{year}{2005}.
\newblock \bibinfo{title}{The fast fourier transform for experimentalists, part
  i: Concepts}.
\newblock \bibinfo{journal}{Computing in Science and Engineering}
  \bibinfo{volume}{7}, \bibinfo{pages}{80--88}.
\newblock \DOIprefix\doi{10.1109/mcse.2005.42}.
\bibitem[{{Event Horizon Telescope Collaboration} et~al.(2019a){Event Horizon
  Telescope Collaboration}, {Akiyama}, {Alberdi}, {Alef}, {Asada}, {Azulay},
  {Baczko}, {Ball}, {Balokovi{\'c}}, {Barrett} and et~al.}]{EHTI}
\bibinfo{author}{{Event Horizon Telescope Collaboration}},
  \bibinfo{author}{{Akiyama}, K.}, \bibinfo{author}{{Alberdi}, A.}, et~al.,
  \bibinfo{year}{2019}a.
\newblock \bibinfo{title}{{First M87 Event Horizon Telescope Results. I. The
  Shadow of the Supermassive Black Hole}}.
\newblock \bibinfo{journal}{\apjl} \bibinfo{volume}{875}, \bibinfo{pages}{L1}.
\newblock \DOIprefix\doi{10.3847/2041-8213/ab0ec7},
  \href{http://arxiv.org/abs/1906.11238}{\tt arXiv:1906.11238}.
\bibitem[{{Event Horizon Telescope Collaboration} et~al.(2019b){Event Horizon
  Telescope Collaboration}, {Akiyama}, {Alberdi}, {Alef}, {Asada}, {Azulay},
  {Baczko}, {Ball}, {Balokovi{\'c}}, {Barrett} and et~al.}]{EHTII}
\bibinfo{author}{{Event Horizon Telescope Collaboration}},
  \bibinfo{author}{{Akiyama}, K.}, \bibinfo{author}{{Alberdi}, A.}, et~al.,
  \bibinfo{year}{2019}b.
\newblock \bibinfo{title}{{First M87 Event Horizon Telescope Results. II. Array
  and Instrumentation}}.
\newblock \bibinfo{journal}{\apjl} \bibinfo{volume}{875}, \bibinfo{pages}{L2}.
\newblock \DOIprefix\doi{10.3847/2041-8213/ab0c96},
  \href{http://arxiv.org/abs/1906.11239}{\tt arXiv:1906.11239}.
\bibitem[{{Event Horizon Telescope Collaboration} et~al.(2019c){Event Horizon
  Telescope Collaboration}, {Akiyama}, {Alberdi}, {Alef}, {Asada}, {Azulay},
  {Baczko}, {Ball}, {Balokovi{\'c}}, {Barrett} and et~al.}]{EHTIII}
\bibinfo{author}{{Event Horizon Telescope Collaboration}},
  \bibinfo{author}{{Akiyama}, K.}, \bibinfo{author}{{Alberdi}, A.}, et~al.,
  \bibinfo{year}{2019}c.
\newblock \bibinfo{title}{{First M87 Event Horizon Telescope Results. III. Data
  Processing and Calibration}}.
\newblock \bibinfo{journal}{\apjl} \bibinfo{volume}{875}, \bibinfo{pages}{L3}.
\newblock \DOIprefix\doi{10.3847/2041-8213/ab0c57},
  \href{http://arxiv.org/abs/1906.11240}{\tt arXiv:1906.11240}.
\bibitem[{{Event Horizon Telescope Collaboration} et~al.(2019d){Event Horizon
  Telescope Collaboration}, {Akiyama}, {Alberdi}, {Alef}, {Asada}, {Azulay},
  {Baczko}, {Ball}, {Balokovi{\'c}}, {Barrett} and et~al.}]{EHTIV}
\bibinfo{author}{{Event Horizon Telescope Collaboration}},
  \bibinfo{author}{{Akiyama}, K.}, \bibinfo{author}{{Alberdi}, A.}, et~al.,
  \bibinfo{year}{2019}d.
\newblock \bibinfo{title}{{First M87 Event Horizon Telescope Results. IV.
  Imaging the Central Supermassive Black Hole}}.
\newblock \bibinfo{journal}{\apjl} \bibinfo{volume}{875}, \bibinfo{pages}{L4}.
\newblock \DOIprefix\doi{10.3847/2041-8213/ab0e85},
  \href{http://arxiv.org/abs/1906.11241}{\tt arXiv:1906.11241}.
\bibitem[{{Event Horizon Telescope Collaboration} et~al.(2019e){Event Horizon
  Telescope Collaboration}, {Akiyama}, {Alberdi}, {Alef}, {Asada}, {Azulay},
  {Baczko}, {Ball}, {Balokovi{\'c}}, {Barrett} and et~al.}]{EHTV}
\bibinfo{author}{{Event Horizon Telescope Collaboration}},
  \bibinfo{author}{{Akiyama}, K.}, \bibinfo{author}{{Alberdi}, A.}, et~al.,
  \bibinfo{year}{2019}e.
\newblock \bibinfo{title}{{First M87 Event Horizon Telescope Results. V.
  Physical Origin of the Asymmetric Ring}}.
\newblock \bibinfo{journal}{\apjl} \bibinfo{volume}{875}, \bibinfo{pages}{L5}.
\newblock \DOIprefix\doi{10.3847/2041-8213/ab0f43},
  \href{http://arxiv.org/abs/1906.11242}{\tt arXiv:1906.11242}.
\bibitem[{{Event Horizon Telescope Collaboration} et~al.(2019f){Event Horizon
  Telescope Collaboration}, {Akiyama}, {Alberdi}, {Alef}, {Asada}, {Azulay},
  {Baczko}, {Ball}, {Balokovi{\'c}}, {Barrett} and et~al.}]{EHTVI}
\bibinfo{author}{{Event Horizon Telescope Collaboration}},
  \bibinfo{author}{{Akiyama}, K.}, \bibinfo{author}{{Alberdi}, A.}, et~al.,
  \bibinfo{year}{2019}f.
\newblock \bibinfo{title}{{First M87 Event Horizon Telescope Results. VI. The
  Shadow and Mass of the Central Black Hole}}.
\newblock \bibinfo{journal}{\apjl} \bibinfo{volume}{875}, \bibinfo{pages}{L6}.
\newblock \DOIprefix\doi{10.3847/2041-8213/ab1141},
  \href{http://arxiv.org/abs/1906.11243}{\tt arXiv:1906.11243}.
\bibitem[{Fish et~al.(2011)Fish, Doeleman, Beaudoin, Blundell, Bolin, Bower,
  Chamberlin, Freund, Friberg, Gurwell, Honma, Inoue, Krichbaum, Lamb, Marrone,
  Moran, Oyama, Plambeck, Primiani, Rogers, Smythe, SooHoo, Strittmatter,
  Tilanus, Titus, Weintroub, Wright, Woody, Young and Ziurys}]{1_3mmVLBI_SgrA}
\bibinfo{author}{Fish, V.L.}, \bibinfo{author}{Doeleman, S.S.},
  \bibinfo{author}{Beaudoin, C.}, et~al., \bibinfo{year}{2011}.
\newblock \bibinfo{title}{1.3 mm wavelength vlbi of {Sagittarius A*}: detection
  of time-variable emission on event horizon scales}.
\newblock \bibinfo{journal}{The Astrophysical Journal} \bibinfo{volume}{727},
  \bibinfo{pages}{L36}.
\newblock \DOIprefix\doi{10.1088/2041-8205/727/2/l36}.
\bibitem[{Fish et~al.(2019)Fish, Shea and Akiyama}]{Fish_2019}
\bibinfo{author}{Fish, V.L.}, \bibinfo{author}{Shea, M.},
  \bibinfo{author}{Akiyama, K.}, \bibinfo{year}{2019}.
\newblock \bibinfo{title}{{Imaging Black Holes and Jets with a VLBI Array
  Including Multiple Space-Based Telescopes}}.
\newblock \bibinfo{journal}{ArXiv e-prints}
  \href{http://arxiv.org/abs/1903.09539}{\tt arXiv:1903.09539}.
\bibitem[{Genzel et~al.(2010)Genzel, Eisenhauer and Gillessen}]{Genzel_2010}
\bibinfo{author}{Genzel, R.}, \bibinfo{author}{Eisenhauer, F.},
  \bibinfo{author}{Gillessen, S.}, \bibinfo{year}{2010}.
\newblock \bibinfo{title}{The galactic center massive black hole and nuclear
  star cluster}.
\newblock \bibinfo{journal}{Reviews of Modern Physics} \bibinfo{volume}{82},
  \bibinfo{pages}{3121--3195}.
\newblock \DOIprefix\doi{10.1103/revmodphys.82.3121}.
\bibitem[{Hezaveh et~al.(2017)Hezaveh, Levasseur and Marshall}]{Hezaveh_2017}
\bibinfo{author}{Hezaveh, Y.D.}, \bibinfo{author}{Levasseur, L.P.},
  \bibinfo{author}{Marshall, P.J.}, \bibinfo{year}{2017}.
\newblock \bibinfo{title}{Fast automated analysis of strong gravitational
  lenses with convolutional neural networks}.
\newblock \bibinfo{journal}{Nature} \bibinfo{volume}{548},
  \bibinfo{pages}{555--557}.
\newblock \DOIprefix\doi{10.1038/nature23463}.
\bibitem[{{H{\"o}gbom}(1974)}]{CLEAN-1}
\bibinfo{author}{{H{\"o}gbom}, J.A.}, \bibinfo{year}{1974}.
\newblock \bibinfo{title}{{Aperture Synthesis with a Non-Regular Distribution
  of Interferometer Baselines}}.
\newblock \bibinfo{journal}{\aaps} \bibinfo{volume}{15}, \bibinfo{pages}{417}.
\bibitem[{Hong et~al.(2014)Hong, Shen, An and Liu}]{Hong_2014}
\bibinfo{author}{Hong, X.}, \bibinfo{author}{Shen, Z.}, \bibinfo{author}{An,
  T.}, et~al., \bibinfo{year}{2014}.
\newblock \bibinfo{title}{The chinese space millimeter-wavelength {VLBI}
  array{\textemdash}a step toward imaging the most compact astronomical
  objects}.
\newblock \bibinfo{journal}{Acta Astronautica} \bibinfo{volume}{102},
  \bibinfo{pages}{217--225}.
\newblock \DOIprefix\doi{10.1016/j.actaastro.2014.05.026}.
\bibitem[{Hunter(2007)}]{Hunter:2007}
\bibinfo{author}{Hunter, J.D.}, \bibinfo{year}{2007}.
\newblock \bibinfo{title}{Matplotlib: A 2d graphics environment}.
\newblock \bibinfo{journal}{Computing In Science \& Engineering}
  \bibinfo{volume}{9}, \bibinfo{pages}{90--95}.
\bibitem[{Issaoun et~al.(2019)Issaoun, Johnson, Blackburn, Brinkerink,
  Mo{\'{s}}cibrodzka, Chael, Goddi, Mart{\'{\i}}-Vidal, Wagner, Doeleman,
  Falcke, Krichbaum, Akiyama, Bach, Bouman, Bower, Broderick, Cho, Crew,
  Dexter, Fish, Gold, G{\'{o}}mez, Hada, Hern{\'{a}}ndez-G{\'{o}}mez,
  Jan{\ss}en, Kino, Kramer, Loinard, Lu, Markoff, Marrone, Matthews, Moran,
  Müller, Roelofs, Ros, Rottmann, Sanchez, Tilanus, de~Vicente, Wielgus,
  Zensus and Zhao}]{Issaoun_2019}
\bibinfo{author}{Issaoun, S.}, \bibinfo{author}{Johnson, M.D.},
  \bibinfo{author}{Blackburn, L.}, et~al., \bibinfo{year}{2019}.
\newblock \bibinfo{title}{The size, shape, and scattering of {Sagittarius A*}
  at 86 {GHz}: First {VLBI} with {ALMA}}.
\newblock \bibinfo{journal}{The Astrophysical Journal} \bibinfo{volume}{871},
  \bibinfo{pages}{30}.
\newblock \DOIprefix\doi{10.3847/1538-4357/aaf732}.
\bibitem[{Ivanov et~al.(2019)Ivanov, Mikheeva, Lukash, Malinovsky, Chernov,
  Andrianov, Kostenko and Likhachev}]{Ivanov_2019}
\bibinfo{author}{Ivanov, P.B.}, \bibinfo{author}{Mikheeva, E.V.},
  \bibinfo{author}{Lukash, V.N.}, et~al., \bibinfo{year}{2019}.
\newblock \bibinfo{title}{Interferometric observations of supermassive black
  holes in millimeter spectrum band}.
\newblock \bibinfo{journal}{Physics-Uspekhi} \bibinfo{volume}{62},
  \bibinfo{pages}{423--449}.
\newblock \DOIprefix\doi{10.3367/ufne.2018.03.038308}.
\bibitem[{Jacobs et~al.(2017)Jacobs, Glazebrook, Collett, More and
  McCarthy}]{Jacobs_2017}
\bibinfo{author}{Jacobs, C.}, \bibinfo{author}{Glazebrook, K.},
  \bibinfo{author}{Collett, T.}, et~al., \bibinfo{year}{2017}.
\newblock \bibinfo{title}{Finding strong lenses in {CFHTLS} using convolutional
  neural networks}.
\newblock \bibinfo{journal}{Monthly Notices of the Royal Astronomical Society}
  \bibinfo{volume}{471}, \bibinfo{pages}{167--181}.
\newblock \DOIprefix\doi{10.1093/mnras/stx1492}.
\bibitem[{Jones et~al.(2001)Jones, Oliphant, Peterson
  et~al.}]{jones_scipy_2001}
\bibinfo{author}{Jones, E.}, \bibinfo{author}{Oliphant, T.},
  \bibinfo{author}{Peterson, P.}, et~al., \bibinfo{year}{2001}.
\newblock \bibinfo{title}{{SciPy}: Open source scientific tools for python}.
\newblock \URLprefix \url{http://www.scipy.org/}.
\bibitem[{Kamruddin and Dexter(2013)}]{Kamruddin_2013}
\bibinfo{author}{Kamruddin, A.B.}, \bibinfo{author}{Dexter, J.},
  \bibinfo{year}{2013}.
\newblock \bibinfo{title}{A geometric crescent model for black hole images}.
\newblock \bibinfo{journal}{Monthly Notices of the Royal Astronomical Society}
  \bibinfo{volume}{434}, \bibinfo{pages}{765--771}.
\newblock \DOIprefix\doi{10.1093/mnras/stt1068}.
\bibitem[{Kardashev et~al.(2014)Kardashev, Novikov, Lukash, Pilipenko,
  Mikheeva, Bisikalo, Wiebe, Doroshkevich, Zasov, Zinchenko, Ivanov, Kostenko,
  Larchenkova, Likhachev, Malov, Malofeev, Pozanenko, Smirnov, Sobolev,
  Cherepashchuk and Shchekinov}]{Kardashev_2014}
\bibinfo{author}{Kardashev, N.S.}, \bibinfo{author}{Novikov, I.D.},
  \bibinfo{author}{Lukash, V.N.}, et~al., \bibinfo{year}{2014}.
\newblock \bibinfo{title}{Review of scientific topics for the millimetron space
  observatory}.
\newblock \bibinfo{journal}{Physics-Uspekhi} \bibinfo{volume}{57},
  \bibinfo{pages}{1199--1228}.
\newblock \DOIprefix\doi{10.3367/ufne.0184.201412c.1319}.
\bibitem[{Kormendy and Ho(2013)}]{BHmass_determination}
\bibinfo{author}{Kormendy, J.}, \bibinfo{author}{Ho, L.C.},
  \bibinfo{year}{2013}.
\newblock \bibinfo{title}{Coevolution (or not) of supermassive black holes and
  host galaxies}.
\newblock \bibinfo{journal}{Annual Review of Astronomy and Astrophysics}
  \bibinfo{volume}{51}, \bibinfo{pages}{511--653}.
\newblock \DOIprefix\doi{10.1146/annurev-astro-082708-101811}.
\bibitem[{Lanusse et~al.(2017)Lanusse, Ma, Li, Collett, Li, Ravanbakhsh,
  Mandelbaum and P{\'{o}}czos}]{Lanusse_2017}
\bibinfo{author}{Lanusse, F.}, \bibinfo{author}{Ma, Q.}, \bibinfo{author}{Li,
  N.}, et~al., \bibinfo{year}{2017}.
\newblock \bibinfo{title}{{CMU} {DeepLens}: deep learning for automatic
  image-based galaxy{\textendash}galaxy strong lens finding}.
\newblock \bibinfo{journal}{Monthly Notices of the Royal Astronomical Society}
  \bibinfo{volume}{473}, \bibinfo{pages}{3895--3906}.
\newblock \DOIprefix\doi{10.1093/mnras/stx1665}.
\bibitem[{Levasseur et~al.(2017)Levasseur, Hezaveh and
  Wechsler}]{Levasseur_2017}
\bibinfo{author}{Levasseur, L.P.}, \bibinfo{author}{Hezaveh, Y.D.},
  \bibinfo{author}{Wechsler, R.H.}, \bibinfo{year}{2017}.
\newblock \bibinfo{title}{Uncertainties in parameters estimated with neural
  networks: application to strong gravitational lensing}.
\newblock \bibinfo{journal}{The Astrophysical Journal} \bibinfo{volume}{850},
  \bibinfo{pages}{L7}.
\newblock \DOIprefix\doi{10.3847/2041-8213/aa9704}.
\bibitem[{L{\o{}}nning et~al.(2019)L{\o{}}nning, Putzky, Sonke, Reneman, Caan
  and Welling}]{Loenning_2019}
\bibinfo{author}{L{\o{}}nning, K.}, \bibinfo{author}{Putzky, P.},
  \bibinfo{author}{Sonke, J.J.}, et~al., \bibinfo{year}{2019}.
\newblock \bibinfo{title}{Recurrent inference machines for reconstructing
  heterogeneous mri data}.
\newblock \bibinfo{journal}{Medical Image Analysis} \bibinfo{volume}{53},
  \bibinfo{pages}{64--78}.
\newblock \DOIprefix\doi{10.1016/j.media.2019.01.005}.
\bibitem[{Lu et~al.(2018)Lu, Krichbaum, Roy, Fish, Doeleman, Johnson, Akiyama,
  Psaltis, Alef, Asada, Beaudoin, Bertarini, Blackburn, Blundell, Bower,
  Brinkerink, Broderick, Cappallo, Crew, Dexter, Dexter, Falcke, Freund,
  Friberg, Greer, Gurwell, Ho, Honma, Inoue, Kim, Lamb, Lindqvist, Macmahon,
  Marrone, Mart{\'{\i}}-Vidal, Menten, Moran, Nagar, Plambeck, Primiani,
  Rogers, Ros, Rottmann, SooHoo, Spilker, Stone, Strittmatter, Tilanus, Titus,
  Vertatschitsch, Wagner, Weintroub, Wright, Young, Zensus and
  Ziurys}]{Lu_2018}
\bibinfo{author}{Lu, R.S.}, \bibinfo{author}{Krichbaum, T.P.},
  \bibinfo{author}{Roy, A.L.}, et~al., \bibinfo{year}{2018}.
\newblock \bibinfo{title}{Detection of intrinsic source structure at $\sim$3
  schwarzschild radii with millimeter-{VLBI} observations of {Sagittarius A*}}.
\newblock \bibinfo{journal}{The Astrophysical Journal} \bibinfo{volume}{859},
  \bibinfo{pages}{60}.
\newblock \DOIprefix\doi{10.3847/1538-4357/aabe2e}.
\bibitem[{{Luminet}(1979)}]{Luminet}
\bibinfo{author}{{Luminet}, J.P.}, \bibinfo{year}{1979}.
\newblock \bibinfo{title}{{Image of a spherical black hole with thin accretion
  disk}}.
\newblock \bibinfo{journal}{\aap} \bibinfo{volume}{75},
  \bibinfo{pages}{228--235}.
\bibitem[{{Luminet}(2019)}]{luminet_review}
\bibinfo{author}{{Luminet}, J.P.}, \bibinfo{year}{2019}.
\newblock \bibinfo{title}{{An Illustrated History of Black Hole Imaging :
  Personal Recollections (1972-2002)}}.
\newblock \bibinfo{journal}{ArXiv e-prints}
  \href{http://arxiv.org/abs/1902.11196}{\tt arXiv:1902.11196}.
\bibitem[{McConnell et~al.(2012)McConnell, Ma, Murphy, Gebhardt, Lauer, Graham,
  Wright and Richstone}]{McConnell_2012}
\bibinfo{author}{McConnell, N.J.}, \bibinfo{author}{Ma, C.P.},
  \bibinfo{author}{Murphy, J.D.}, et~al., \bibinfo{year}{2012}.
\newblock \bibinfo{title}{Dynamical measurements of black hole masses in four
  brightest cluster galaxies at 100~mpc}.
\newblock \bibinfo{journal}{The Astrophysical Journal} \bibinfo{volume}{756},
  \bibinfo{pages}{179}.
\newblock \DOIprefix\doi{10.1088/0004-637x/756/2/179}.
\bibitem[{Michelson and Pease(1921)}]{Michelson_1921}
\bibinfo{author}{Michelson, A.A.}, \bibinfo{author}{Pease, F.G.},
  \bibinfo{year}{1921}.
\newblock \bibinfo{title}{Measurement of the diameter of alpha orionis with the
  interferometer.}
\newblock \bibinfo{journal}{The Astrophysical Journal} \bibinfo{volume}{53},
  \bibinfo{pages}{249}.
\newblock \URLprefix \url{https://doi.org/10.1086%2F142603},
  \DOIprefix\doi{10.1086/142603}.
\bibitem[{{Morningstar} et~al.(2018){Morningstar}, {Hezaveh}, {Perreault
  Levasseur}, {Blandford}, {Marshall}, {Putzky} and
  {Wechsler}}]{Morningstar_2018}
\bibinfo{author}{{Morningstar}, W.R.}, \bibinfo{author}{{Hezaveh}, Y.D.},
  \bibinfo{author}{{Perreault Levasseur}, L.}, et~al., \bibinfo{year}{2018}.
\newblock \bibinfo{title}{{Analyzing interferometric observations of strong
  gravitational lenses with recurrent and convolutional neural networks}}.
\newblock \bibinfo{journal}{arXiv e-prints} ,
  \bibinfo{pages}{arXiv:1808.00011}\href{http://arxiv.org/abs/1808.00011}{\tt
  arXiv:1808.00011}.
\bibitem[{Morningstar et~al.(2019)Morningstar, Levasseur, Hezaveh, Blandford,
  Marshall, Putzky, Rueter, Wechsler and Welling}]{Morningstar_2019}
\bibinfo{author}{Morningstar, W.R.}, \bibinfo{author}{Levasseur, L.P.},
  \bibinfo{author}{Hezaveh, Y.D.}, et~al., \bibinfo{year}{2019}.
\newblock \bibinfo{title}{Data-driven reconstruction of gravitationally lensed
  galaxies using recurrent inference machines}.
\newblock \bibinfo{journal}{The Astrophysical Journal} \bibinfo{volume}{883},
  \bibinfo{pages}{14}.
\newblock \URLprefix \url{https://doi.org/10.3847%2F1538-4357%2Fab35d7},
  \DOIprefix\doi{10.3847/1538-4357/ab35d7},
  \href{http://arxiv.org/abs/1901.01359}{\tt arXiv:1901.01359}.
\bibitem[{Narayan et~al.(2019)Narayan, Johnson and Gammie}]{Narayan_2019}
\bibinfo{author}{Narayan, R.}, \bibinfo{author}{Johnson, M.D.},
  \bibinfo{author}{Gammie, C.F.}, \bibinfo{year}{2019}.
\newblock \bibinfo{title}{The shadow of a spherically accreting black hole}.
\newblock \bibinfo{journal}{The Astrophysical Journal} \bibinfo{volume}{885},
  \bibinfo{pages}{L33}.
\newblock \URLprefix \url{https://doi.org/10.3847%2F2041-8213%2Fab518c},
  \DOIprefix\doi{10.3847/2041-8213/ab518c}.
\bibitem[{Narayan and Nityananda(1986)}]{Narayan_1986}
\bibinfo{author}{Narayan, R.}, \bibinfo{author}{Nityananda, R.},
  \bibinfo{year}{1986}.
\newblock \bibinfo{title}{Maximum entropy image restoration in astronomy}.
\newblock \bibinfo{journal}{Annual Review of Astronomy and Astrophysics}
  \bibinfo{volume}{24}, \bibinfo{pages}{127--170}.
\newblock \DOIprefix\doi{10.1146/annurev.aa.24.090186.001015}.
\bibitem[{{Palumbo} et~al.(2019){Palumbo}, {Doeleman}, {Johnson}
  et~al.}]{Palumbo}
\bibinfo{author}{{Palumbo}, D.C.M.}, \bibinfo{author}{{Doeleman}, S.S.},
  \bibinfo{author}{{Johnson}, M.D.}, et~al., \bibinfo{year}{2019}.
\newblock \bibinfo{title}{{Metrics and Motivations for Earth-Space VLBI:
  Time-Resolving Sgr A* with the Event Horizon Telescope}}.
\newblock \bibinfo{journal}{ArXiv e-prints}
  \href{http://arxiv.org/abs/1906.08828}{\tt arXiv:1906.08828}.
\bibitem[{P\'erez and Granger(2007)}]{PER-GRA:2007}
\bibinfo{author}{P\'erez, F.}, \bibinfo{author}{Granger, B.E.},
  \bibinfo{year}{2007}.
\newblock \bibinfo{title}{{IP}ython: a system for interactive scientific
  computing}.
\newblock \bibinfo{journal}{Computing in Science and Engineering}
  \bibinfo{volume}{9}, \bibinfo{pages}{21--29}.
\newblock \URLprefix \url{http://ipython.org},
  \DOIprefix\doi{10.1109/MCSE.2007.53}.
\bibitem[{Petrillo et~al.(2017)Petrillo, Tortora, Chatterjee, Vernardos,
  Koopmans, Kleijn, Napolitano, Covone, Schneider, Grado and
  McFarland}]{Petrillo_2017}
\bibinfo{author}{Petrillo, C.E.}, \bibinfo{author}{Tortora, C.},
  \bibinfo{author}{Chatterjee, S.}, et~al., \bibinfo{year}{2017}.
\newblock \bibinfo{title}{Finding strong gravitational lenses in the {Kilo
  Degree Survey} with {Convolutional Neural Networks}}.
\newblock \bibinfo{journal}{Monthly Notices of the Royal Astronomical Society}
  \bibinfo{volume}{472}, \bibinfo{pages}{1129--1150}.
\newblock \DOIprefix\doi{10.1093/mnras/stx2052}.
\bibitem[{{Popov} et~al.(2021){Popov}, {Strokov} and
  {Surdyaev}}]{2021A&C....3600467P}
\bibinfo{author}{{Popov}, A.A.}, \bibinfo{author}{{Strokov}, V.N.},
  \bibinfo{author}{{Surdyaev}, A.A.}, \bibinfo{year}{2021}.
\newblock \bibinfo{title}{{A proof-of-concept neural network for inferring
  parameters of a black hole from partial interferometric images of its
  shadow}}.
\newblock \bibinfo{journal}{Astronomy and Computing} \bibinfo{volume}{36},
  \bibinfo{pages}{100467}.
\newblock \DOIprefix\doi{10.1016/j.ascom.2021.100467}.
\bibitem[{{Porth} et~al.(2019){Porth}, {Chatterjee}, {Narayan}, {Gammie},
  {Mizuno}, {Anninos}, {Baker}, {Bugli}, {Chan}, {Davelaar}, {Del Zanna},
  {Etienne}, {Fragile}, {Kelly}, {Liska}, {Markoff}, {McKinney}, {Mishra},
  {Noble}, {Olivares}, {Prather}, {Rezzolla}, {Ryan}, {Stone}, {Tomei},
  {White}, {Younsi}, {Akiyama}, {Alberdi}, {Alef}, {Asada}, {Azulay}, {Baczko},
  {Ball}, {Balokovi{\'c}}, {Barrett}, {Bintley}, {Blackburn}, {Boland},
  {Bouman}, {Bower}, {Bremer}, {Brinkerink}, {Brissenden}, {Britzen},
  {Broderick}, {Broguiere}, {Bronzwaer}, {Byun}, {Carlstrom}, {Chael},
  {Chatterjee}, {Chen}, {Chen}, {Cho}, {Christian}, {Conway}, {Cordes},
  {Geoffrey}, {Crew}, {Cui}, {De Laurentis}, {Deane}, {Dempsey}, {Desvignes},
  {Doeleman}, {Eatough}, {Falcke}, {Fish}, {Fomalont}, {Fraga-Encinas},
  {Freeman}, {Friberg}, {Fromm}, {G{\'o}mez}, {Galison}, {Garc{\'\i}a},
  {Gentaz}, {Georgiev}, {Goddi}, {Gold}, {Gu}, {Gurwell}, {Hada}, {Hecht},
  {Hesper}, {Ho}, {Ho}, {Honma}, {Huang}, {Huang}, {Hughes}, {Ikeda}, {Inoue},
  {Issaoun}, {James}, {Jannuzi}, {Janssen}, {Jeter}, {Jiang}, {Johnson},
  {Jorstad}, {Jung}, {Karami}, {Karuppusamy}, {Kawashima}, {Keating},
  {Kettenis}, {Kim}, {Kim}, {Kim}, {Kino}, {Koay}, {Patrick}, {Koch}, {Koyama},
  {Kramer}, {Kramer}, {Krichbaum}, {Kuo}, {Lauer}, {Lee}, {Li}, {Li},
  {Lindqvist}, {Liu}, {Liuzzo}, {Lo}, {Lobanov}, {Loinard}, {Lonsdale}, {Lu},
  {MacDonald}, {Mao}, {Marrone}, {Marscher}, {Mart{\'\i}-Vidal}, {Matsushita},
  {Matthews}, {Medeiros}, {Menten}, {Mizuno}, {Moran}, {Moriyama},
  {Moscibrodzka}, {M{\"u}ller}, {Nagai}, {Nagar}, {Nakamura}, {Narayanan},
  {Natarajan}, {Neri}, {Ni}, {Noutsos}, {Okino}, {Oyama}, {{\"O}zel},
  {Palumbo}, {Patel}, {Pen}, {Pesce}, {Pi{\'e}tu}, {Plambeck}, {PopStefanija},
  {Preciado-L{\'o}pez}, {Psaltis}, {Pu}, {Ramakrishnan}, {Rao}, {Rawlings},
  {Raymond}, {Ripperda}, {Roelofs}, {Rogers}, {Ros}, {Rose}, {Roshanineshat},
  {Rottmann}, {Roy}, {Ruszczyk}, {Rygl}, {S{\'a}nchez},
  {S{\'a}nchez-Arguelles}, {Sasada}, {Savolainen}, {Schloerb}, {Schuster},
  {Shao}, {Shen}, {Small}, {Sohn}, {SooHoo}, {Tazaki}, {Tiede}, {Tilanus},
  {Titus}, {Toma}, {Torne}, {Trent}, {Trippe}, {Tsuda}, {van Bemmel}, {van
  Langevelde}, {van Rossum}, {Wagner}, {Wardle}, {Weintroub}, {Wex}, {Wharton},
  {Wielgus}, {Wong}, {Wu}, {Young}, {Young}, {Yuan}, {Yuan}, {Zensus}, {Zhao},
  {Zhao}, {Zhu} and {Event Horizon Telescope Collaboration}}]{harms}
\bibinfo{author}{{Porth}, O.}, \bibinfo{author}{{Chatterjee}, K.},
  \bibinfo{author}{{Narayan}, R.}, et~al., \bibinfo{year}{2019}.
\newblock \bibinfo{title}{{The Event Horizon General Relativistic
  Magnetohydrodynamic Code Comparison Project}}.
\newblock \bibinfo{journal}{\apjs} \bibinfo{volume}{243}, \bibinfo{pages}{26}.
\newblock \DOIprefix\doi{10.3847/1538-4365/ab29fd},
  \href{http://arxiv.org/abs/1904.04923}{\tt arXiv:1904.04923}.
\bibitem[{Pourrahmani et~al.(2018)Pourrahmani, Nayyeri and
  Cooray}]{Pourrahmani_2018}
\bibinfo{author}{Pourrahmani, M.}, \bibinfo{author}{Nayyeri, H.},
  \bibinfo{author}{Cooray, A.}, \bibinfo{year}{2018}.
\newblock \bibinfo{title}{{LensFlow}: A convolutional neural network in search
  of strong gravitational lenses}.
\newblock \bibinfo{journal}{The Astrophysical Journal} \bibinfo{volume}{856},
  \bibinfo{pages}{68}.
\newblock \URLprefix \url{https://doi.org/10.3847%2F1538-4357%2Faaae6a},
  \DOIprefix\doi{10.3847/1538-4357/aaae6a},
  \href{http://arxiv.org/abs/1705.05857}{\tt arXiv:1705.05857}.
\bibitem[{{Press} et~al.(2002){Press}, {Teukolsky}, {Vetterling} and
  {Flannery}}]{recipes}
\bibinfo{author}{{Press}, W.H.}, \bibinfo{author}{{Teukolsky}, S.A.},
  \bibinfo{author}{{Vetterling}, W.T.}, et~al., \bibinfo{year}{2002}.
\newblock \bibinfo{title}{{Numerical recipes in C++ : the art of scientific
  computing}}.
\bibitem[{{Putzky} and {Welling}(2017)}]{Putzky_2017}
\bibinfo{author}{{Putzky}, P.}, \bibinfo{author}{{Welling}, M.},
  \bibinfo{year}{2017}.
\newblock \bibinfo{title}{{Recurrent Inference Machines for Solving Inverse
  Problems}}.
\newblock \bibinfo{journal}{arXiv e-prints} ,
  \bibinfo{pages}{arXiv:1706.04008}\href{http://arxiv.org/abs/1706.04008}{\tt
  arXiv:1706.04008}.
\bibitem[{{Ruprecht} et~al.(2011){Ruprecht}, {Johannsen}, {Fish}, {Broderick},
  {Doeleman}, {Loeb} and {Rogers}}]{EHT-3}
\bibinfo{author}{{Ruprecht}, J.}, \bibinfo{author}{{Johannsen}, T.},
  \bibinfo{author}{{Fish}, V.L.}, et~al., \bibinfo{year}{2011}.
\newblock \bibinfo{title}{{Testing general relativity with the Event Horizon
  Telescope}}, in: \bibinfo{booktitle}{American Astronomical Society Meeting
  Abstracts \#218}, p. \bibinfo{pages}{229.07}.
\bibitem[{{Russakovsky} et~al.(2014){Russakovsky}, {Deng}, {Su}, {Krause},
  {Satheesh}, {Ma}, {Huang}, {Karpathy}, {Khosla}, {Bernstein}, {Berg} and
  {Fei-Fei}}]{russakovsky}
\bibinfo{author}{{Russakovsky}, O.}, \bibinfo{author}{{Deng}, J.},
  \bibinfo{author}{{Su}, H.}, et~al., \bibinfo{year}{2014}.
\newblock \bibinfo{title}{{ImageNet Large Scale Visual Recognition Challenge}}.
\newblock \bibinfo{journal}{arXiv e-prints}
  \href{http://arxiv.org/abs/1409.0575}{\tt arXiv:1409.0575}.
\bibitem[{Schaefer et~al.(2018)Schaefer, Geiger, Kuntzer and
  Kneib}]{Schaefer_2018}
\bibinfo{author}{Schaefer, C.}, \bibinfo{author}{Geiger, M.},
  \bibinfo{author}{Kuntzer, T.}, et~al., \bibinfo{year}{2018}.
\newblock \bibinfo{title}{Deep convolutional neural networks as strong
  gravitational lens detectors}.
\newblock \bibinfo{journal}{Astronomy {\&} Astrophysics} \bibinfo{volume}{611},
  \bibinfo{pages}{A2}.
\newblock \URLprefix \url{https://doi.org/10.1051%2F0004-6361%2F201731201},
  \DOIprefix\doi{10.1051/0004-6361/201731201},
  \href{http://arxiv.org/abs/1705.07132}{\tt arXiv:1705.07132}.
\bibitem[{Shapiro and Wilk(1965)}]{SHAPIRO_1965}
\bibinfo{author}{Shapiro, S.S.}, \bibinfo{author}{Wilk, M.B.},
  \bibinfo{year}{1965}.
\newblock \bibinfo{title}{An analysis of variance test for normality (complete
  samples)}.
\newblock \bibinfo{journal}{Biometrika} \bibinfo{volume}{52},
  \bibinfo{pages}{591--611}.
\newblock \DOIprefix\doi{10.1093/biomet/52.3-4.591}.
\bibitem[{{The Theano Development Team} et~al.(2016){The Theano Development
  Team}, {Al-Rfou}, {Alain}, {Almahairi}, {Angermueller}, {Bahdanau}, {Ballas},
  {Bastien}, {Bayer}, {Belikov}, {Belopolsky}, {Bengio}, {Bergeron},
  {Bergstra}, {Bisson}, {Bleecher Snyder}, {Bouchard}, {Boulanger-Lewandowski},
  {Bouthillier}, {de Br{\'e}bisson}, {Breuleux}, {Carrier}, {Cho}, {Chorowski},
  {Christiano}, {Cooijmans}, {C{\^o}t{\'e}}, {C{\^o}t{\'e}}, {Courville},
  {Dauphin}, {Delalleau}, {Demouth}, {Desjardins}, {Dieleman}, {Dinh},
  {Ducoffe}, {Dumoulin}, {Ebrahimi Kahou}, {Erhan}, {Fan}, {Firat}, {Germain},
  {Glorot}, {Goodfellow}, {Graham}, {Gulcehre}, {Hamel}, {Harlouchet}, {Heng},
  {Hidasi}, {Honari}, {Jain}, {Jean}, {Jia}, {Korobov}, {Kulkarni}, {Lamb},
  {Lamblin}, {Larsen}, {Laurent}, {Lee}, {Lefrancois}, {Lemieux},
  {L{\'e}onard}, {Lin}, {Livezey}, {Lorenz}, {Lowin}, {Ma}, {Manzagol},
  {Mastropietro}, {McGibbon}, {Memisevic}, {van Merri{\"e}nboer}, {Michalski},
  {Mirza}, {Orlandi}, {Pal}, {Pascanu}, {Pezeshki}, {Raffel}, {Renshaw},
  {Rocklin}, {Romero}, {Roth}, {Sadowski}, {Salvatier}, {Savard},
  {Schl{\"u}ter}, {Schulman}, {Schwartz}, {Vlad Serban}, {Serdyuk},
  {Shabanian}, {Simon}, {Spieckermann}, {Ramana Subramanyam}, {Sygnowski},
  {Tanguay}, {van Tulder}, {Turian}, {Urban}, {Vincent}, {Visin}, {de Vries},
  {Warde-Farley}, {Webb}, {Willson}, {Xu}, {Xue}, {Yao}, {Zhang} and
  {Zhang}}]{theano}
\bibinfo{author}{{The Theano Development Team}}, \bibinfo{author}{{Al-Rfou},
  R.}, \bibinfo{author}{{Alain}, G.}, et~al., \bibinfo{year}{2016}.
\newblock \bibinfo{title}{{Theano: A Python framework for fast computation of
  mathematical expressions}}.
\newblock \bibinfo{journal}{arXiv e-prints} ,
  \bibinfo{pages}{arXiv:1605.02688}\href{http://arxiv.org/abs/1605.02688}{\tt
  arXiv:1605.02688}.
\bibitem[{Thompson et~al.(2017)Thompson, Moran and Swenson}]{Thompson_2017}
\bibinfo{author}{Thompson, A.R.}, \bibinfo{author}{Moran, J.M.},
  \bibinfo{author}{Swenson, G.W.}, \bibinfo{year}{2017}.
\newblock \bibinfo{title}{Interferometry and Synthesis in Radio Astronomy}.
\newblock \bibinfo{publisher}{Springer International Publishing}.
\newblock \DOIprefix\doi{10.1007/978-3-319-44431-4}.
\bibitem[{{van der Gucht} et~al.(2019){van der Gucht}, {Davelaar}, {Hendriks},
  {Porth}, {Olivares}, {Mizuno}, {Fromm} and {Falcke}}]{deep_horizon}
\bibinfo{author}{{van der Gucht}, J.}, \bibinfo{author}{{Davelaar}, J.},
  \bibinfo{author}{{Hendriks}, L.}, et~al., \bibinfo{year}{2019}.
\newblock \bibinfo{title}{{Deep Horizon; a machine learning network that
  recovers accreting black hole parameters}}.
\newblock \bibinfo{journal}{ArXiv e-prints}
  \href{http://arxiv.org/abs/1910.13236}{\tt arXiv:1910.13236}.
\bibitem[{Van Der~Walt et~al.(2011)Van Der~Walt, Colbert and
  Varoquaux}]{van2011numpy}
\bibinfo{author}{Van Der~Walt, S.}, \bibinfo{author}{Colbert, S.C.},
  \bibinfo{author}{Varoquaux, G.}, \bibinfo{year}{2011}.
\newblock \bibinfo{title}{The numpy array: a structure for efficient numerical
  computation}.
\newblock \bibinfo{journal}{Computing in Science \& Engineering}
  \bibinfo{volume}{13}, \bibinfo{pages}{22--30}.

\end{thebibliography}

\end{document}